%% file: poolingLPs.tex
\documentclass[12pt,a4paper,leqno]{article}
\usepackage[utf8]{inputenc}
\usepackage{geometry}
\usepackage{xcolor}
\usepackage[round]{natbib}
\usepackage{graphicx}
\usepackage{pdflscape}
\usepackage{appendix}
\usepackage{threeparttable}
\usepackage{booktabs}
\usepackage{longtable}

\usepackage{amsthm}

\usepackage{amssymb}
\usepackage{amsbsy}
\usepackage{bm}
\usepackage{amsmath}
\usepackage{amsthm}
\usepackage{graphicx}
\usepackage{mathtools}
\usepackage{rotating}
\usepackage{setspace}
\usepackage{footmisc}
\usepackage{color, colortbl}
\definecolor{ashgrey}{rgb}{0.7, 0.75, 0.71}
\definecolor{columbiablue}{rgb}{0.61, 0.87, 1.0}
\definecolor{coral}{rgb}{1.0, 0.5, 0.31}

\definecolor{colBVAR}{HTML}{bababa}
\definecolor{colBART}{HTML}{d7191c}
\definecolor{colmixBART}{HTML}{fdae61}
\definecolor{colerrorBART}{HTML}{abd9e9}
\definecolor{colfullBART}{HTML}{2c7bb6}

\definecolor{colcons}{HTML}{e31a1c}
\definecolor{colSV}{HTML}{a6cee3}
\definecolor{colhBART}{HTML}{1f78b4}

\usepackage{xcolor}
\usepackage{tikz}

\usepackage{enumitem}
\newlist{steps}{enumerate}{1}
\setlist[steps,1]{label = Step \arabic*:}

\usepackage{adjustbox}
\usepackage{dcolumn}
\newcolumntype{d}[1]{D..{#1}} 

\usepackage{caption}
\usepackage{subcaption}
\definecolor{nblue}{HTML}{000660}
\usepackage[colorlinks=true,urlcolor=nblue,linkcolor=nblue,citecolor=nblue]{hyperref}
\newcommand{\N}{\mathcal{N}}

\graphicspath{{figures/}}

\newcommand{\fignotes}[1]{\par\vspace{4pt}\begin{minipage}{\linewidth}\tiny\emph{Notes:} #1\end{minipage}}

\usepackage[compact]{titlesec}
\titlespacing*{\section}{0pt}{1.2ex plus 0.4ex}{0.8ex plus 0.2ex}
\titlespacing*{\subsection}{0pt}{1.0ex plus 0.3ex}{0.6ex plus 0.2ex}

\begin{document}

\title{\textbf{Clustered Local Projections for Short and Ultra-Short Time Series\\
\large A Hierarchical Bayesian Framework}\thanks{The views expressed herein are solely those of the authors and do not necessarily reflect the views of the Federal Reserve Bank of Cleveland or the Federal Reserve System.  We acknowledge the use of Claude Opus as an aid for coding and for drafting, editing, and brainstorming during the preparation of this manuscript. The authors acknowledge the computational resources and services provided by Salzburg Collaborative Computing (SCC), funded by the Federal Ministry of Education, Science and Research (BMBWF) and the State of Salzburg. Huber gratefully acknowledges funding from the Jubil\"{a}umsfonds of the Oesterreichische Nationalbank (OeNB): 19090.}}

\author{Todd E.\ Clark\thanks{Economist Emeritus, Federal Reserve Bank of Cleveland, and Fellow, Dept.\ of Economics, Johns Hopkins University} \and
Florian Huber\thanks{Professor of Economics, University of Salzburg}}

\date{}

\maketitle
\thispagestyle{empty}

\doublespacing
\begin{center}
\begin{minipage}{0.9\textwidth}
\noindent\small
\begin{center}
    \textbf{Abstract}\\
\end{center}
Estimating the dynamic effects of economic shocks in short and very short samples is impeded by a lack of degrees of freedom. We offer a solution based on a Bayesian hierarchical framework for estimating local projection (LP) impulse response functions across a panel of related time series. The framework explicitly accommodates unbalanced panels in which some series are substantially shorter than others, allowing the short series to borrow information from longer ones at horizons where the short series carry little or no own data. Since series might exhibit heterogeneous dynamics, we develop a sparse finite mixture pool that clusters units by similarity of their impulse response profiles. We show in simulations that our approach substantially improves LP estimation accuracy relative to the standard approach if the time series are short while producing similar LPs for longer time series. Using a US price dataset, augmented with survey responses, we find that supply-chain and oil shocks trigger heterogeneous reactions of different price measures, with headline price indices responding more sharply than their core counterparts and goods prices changing more than services prices.
\end{minipage}
\end{center}

\setcounter{page}{0}

\setlength{\abovedisplayskip}{8pt plus 2pt minus 2pt}
\setlength{\belowdisplayskip}{8pt plus 2pt minus 2pt}
\setlength{\abovedisplayshortskip}{4pt plus 2pt}
\setlength{\belowdisplayshortskip}{5pt plus 2pt minus 2pt}
\normalsize\newpage\renewcommand{\footnotelayout}{\setstretch{1}}

\section{Introduction}

Local projections (LPs), introduced by \citet{jorda2005}, have become a popular approach for estimating the dynamic causal effects of identified structural shocks. Their appeal is easy to state. Each horizon is a separate regression of a future outcome on the shock and a set of controls. No dynamic model needs to be specified and iterated forward, and misspecification at one horizon does not propagate to the others.  This convenience comes at a price: LPs are data hungry. The horizon-$h$ LP sacrifices $h$ observations at the end of the sample, so the effective sample shrinks exactly for the objects of interest, such as the responses at longer horizons. 

In macroeconomics, this problem is intensified due to the relatively short samples we have available. In this case, \citet{HerbstJohannsen2024} document that short samples contribute to bias in LP estimates along with coverage rates that are below nominal levels. Another problem is that many of the newly  available time series  start late, implying that in some cases the horizon $h$ could even exceed the length of this time series.  

One observation, however, is that the series in macroeconomic panels often share similar time series properties. This holds not only for longer time series but also for newly developed indicators that might be related to more established longer time series in the panel.  This observation motivates the present paper. In the applications we have in mind, the short series are related to one another as well as to some of the available long time series. For instance,  a newly introduced services producer price index (PPI) enters a panel alongside dozens of other consumer and producer price series, some reaching back to the late 1950s, whose responses to the same identified shocks are plausibly similar. Our goal is to develop an econometric model that exploits precisely this information to improve LP estimates of short and ultra-short time series.

We propose a Bayesian hierarchical framework for estimating LPs  across  a panel of related time series that may differ in terms of their individual sample sizes. We remain within the linear LP framework but rely on Bayesian estimation with priors that pool information across variables, especially on the response coefficients of interest. Our point of departure  is a simple pooling prior under which the horizon-$h$ response of each series is drawn from a common distribution. The posterior for any given series is then a precision-weighted average of its own least-squares estimate and the panel-wide mean, and the weight on the cross-section rises endogenously as the series' own sample shrinks. Coefficient estimates of long series are close to the corresponding ordinary least squares (OLS) estimates, while short series profit from the information arising from having longer series with similar characteristics in the panel.

Pooling towards a single common response is, of course, the wrong thing to do if the panel is heterogeneous. For instance, there is no reason why gasoline prices and hospital services prices should react alike to a monetary policy shock. We therefore generalize the pooling prior to a sparse finite mixture in the spirit of \citet{malsinerwalli2016}. Our model allocates the series in the panel probabilistically to a small number of latent clusters. A Gaussian pooling prior that is  now centered on cluster-specific rather than panel-wide means  then pulls each series' response towards its own cluster's mean, while the effective number of clusters (the groups of series in the panel that share similar characteristics) is learned from the data through a sparse Dirichlet prior on the mixture weights. A third layer of the hierarchy ties the cluster means to a common population center, so that small clusters, including those with a single member, continue to borrow strength, more weakly, from the panel as a whole. Which series end up being grouped together is itself of economic interest since the clusters are formed on the similarity of impulse response functions (IRFs) rather than on any a priori sectoral classification. Series that appear only loosely related at a first glance can end up informing one another.

A key property of the framework is that pooling series of very different lengths requires neither interpolation nor any balancing of the panel. The reason is that the LP equations are, conditional on the parameters of the pooling prior, independent across series and horizons. A shorter series simply contributes fewer observations to the cross-sectional updating steps, and its own estimates place correspondingly more weight on the pooling prior. If a horizon of interest exceeds a series' effective sample length (as in the ultra-short case), the data contribute nothing at all and the response is a draw from the relevant cluster's distribution. The framework produces IRFs, with all estimation uncertainty propagated, at horizons the series alone could never reach, while for horizons within the sample the estimates remain firmly informed by the series' own data.

An issue that any likelihood-based treatment of LPs must confront is that the horizon-$h$ error inherits a moving average (MA) structure, so that credible sets computed from the equation-by-equation posterior understate uncertainty. Rather than modeling the serial correlation, which would induce correlations among the  LP equations across horizons, we correct for it ex post, following \citet{muller2013}. The posterior draws are rescaled so that the resulting credible sets are calibrated to a sandwich variance, which makes them approximately valid in the frequentist sense. The long-run variance this requires is pooled at the cluster level, over the same partition that pools the IRFs. Pooling the correction matters. Unit-specific long-run variances are extremely noisy in exactly the short samples we care about, and it is the cluster-level version that makes the correction operational in such a setting.

Taken together, the framework has four features that distinguish it from existing work. First,  hierarchical pooling on the IRF coefficients, with the pooling intensity learned from the data. Second,  a sparse finite mixture that groups series by the similarity of their responses, with the effective number of groups determined endogenously. Third,  a three-level hierarchy that pools cluster means towards a population center and thereby handles small clusters efficiently. Fourth, and finally,  asymmetric-panel borrowing that assigns short series estimates at horizons beyond their effective sample length, with all uncertainty propagated and without interpolating any data.

We assess the framework in a simulation study calibrated to US macro data. The gains are substantial. Relative to series-by-series LPs, the full model  that pools all coefficients with a mixture reduces the mean absolute error of the IRF estimates on short series by a factor of about two at short, medium, and long horizons alike, and is the most accurate of the estimators we compare at nearly every horizon, while on long series it stays within roughly five to seven percent of the series-by-series benchmark. Pooling therefore costs little when a series already has ample data of its own. Forcing a single common response instead produces visible bias on exactly those series. This suggests that it is the mixture, not pooling per se, that makes the pooling safe. On the inference side, the coverage of naive LP confidence intervals on short series degrades to roughly two-thirds at long horizons, whereas the cluster-pooled correction restores coverage to the nominal 90 percent and a little beyond. 

We then apply the framework to a panel of $43$ series ($33$ US consumer and producer price indices augmented with $10$ regional Federal Reserve business-survey price measures) and study their responses to two identified structural shocks: a recently developed supply-chain shock \citep{KanzigRaghavan2026} and the Baumeister--Hamilton oil supply shock \citep{BaumeisterHamilton2019}. The panel has the structure the framework is built for. A handful of consumer price series with histories reaching back to 1959 sit alongside producer price series, notably the newer services PPIs, and the survey-based price measures, all with far shorter samples. The mixture sorts the panel into six or seven clusters with distinct response profiles, broadly separating consumer prices, final-demand producer prices, intermediate-demand goods, and the survey-based measures, and sharpens the responses of the newer, short series, which contribute as few as $33$ usable observations at the longest horizon against more than $500$ for the long-running aggregates.  The estimated impulse responses show that the shocks induce larger movements in headline price indices than those omitting food and energy components and larger changes in goods prices than services prices, with the least processed goods moving most.

Our paper connects to several strands of literature. First, a small but growing literature develops Bayesian approaches to LPs \citep{ferreira2025, Schwarzbach2026blp}. \citet{ferreira2025} develop Bayesian LPs for a single unit, including a treatment of the serial correlation in the multi-step error that any likelihood-based approach to LPs must take a stand on. The closest precedent for our work is \citet{Schwarzbach2026blp}, who proposes Bayesian panel LPs with a somewhat different motivation from ours; their framework amounts to a single pooling specification, without the mixture, the third layer of the hierarchy, or the treatment of severely unbalanced panels that are central to our contribution. Second, pooling priors have a longer tradition in multivariate time series settings.  \citet{FruhwirthSchnatterKaufmann2008} propose model-based clustering of many short time series to improve estimation efficiency. Within a vector autoregression (VAR) framework,  \citet{jarocinski2010} uses a pooling prior to compare responses to monetary policy shocks across countries. More recently,  \citet{HKP2023AoAS} pool coefficients in a Bayesian panel VAR, and \citet{MullerWatsonJAE} develop pooling methods for forecasting related time series, including specifications that pool innovation variances. Relative to \citet{HKP2023AoAS},  we move the pooling from the VAR to the LP setting, where the equation-by-equation structure turns from a nuisance into an asset once the panel is unbalanced. Third, a complementary strand improves the efficiency of LPs by exploiting information across horizons rather than across series, either by smoothing the IRF \citep{BarnichonMatthes2018JME} or by estimating the horizons jointly \citep{huber2024gsulp}; combining across-horizon smoothing with our cross-sectional pooling is a natural extension that we leave for future work. A related strand instead relaxes the functional form of the projection itself \citep[see, e.g.,][]{mumtaz2022}.  Finally, our application speaks to the literature on the responses of disaggregated prices to identified shocks \citep{Clark1999RESTAT, AruobaDrechsel2024JME}, which has largely concentrated on consumer prices and the longer-running producer price aggregates. We suspect this is partly because the newer series are hard to handle with standard methods.

The remainder of the paper is organized as follows. Section~\ref{sec:framework} develops the econometric framework: the hierarchical pooling prior, its sparse finite mixture extension, a sketch of the posterior simulator, and the robustification of the error variances. Section~\ref{sec:simulation} presents the simulation evidence. Section~\ref{sec:application} contains the application to consumer and producer prices. Technical details on the posterior simulator are collected in the Appendix.

\section{Econometric framework}\label{sec:framework}
\subsection{General setup and notation}
We observe $M$ time series $y_{i, t}$ on a common calendar clock $t = 1, \ldots, T$, and we let $\mathcal{O}_i = \{t: y_{i,t} \neq \text{NA}\}$ denote the points in time for which series $i$ has observations.\footnote{Throughout we assume each $\mathcal{O}_i$ is contiguous. Series may start late or end early, but they have no internal gaps, so the series-specific length is $T_i = |\mathcal{O}_i|$. This is the case in our application and it keeps the counting below simple. The hierarchy does not require it. With internal gaps, every expression still holds once the range $t = p+1, \ldots, T_i - h$ is replaced by the set of usable dates and $T_{i,h}$ by the number of such dates.}  We assume that the series share a common identified structural shock $w_t$, observed for $t = 1, \ldots, T$, where $T = \max_i T_i = \max \{T_1, \ldots, T_M\}$.  Let $\bm x_t$ denote a vector of controls (a constant, lags of $y_{i,t}$, lags of $w_t$, and/or additional exogenous covariates) of dimension $k$.

Define the maximum effective horizon for series $i$ as:
\begin{equation*}
H_i = T_i - p - 1,
\end{equation*}
where $p$ is the lag length used in $\bm x_t$. We are interested in horizons $h \in \{0, 1, \ldots, H\}$ where $H \leq \max_i H_i$ so that the longest horizon we consider does not exceed the maximum effective horizon in the panel.

We consider LPs for series $i$ at horizon $h$:
\begin{equation}
y_{i,t+h} = \rho_{i,h} w_t + \bm{\beta}_{i,h}' \bm{x}_t + u_{i,t+h}, \qquad u_{i,t+h} \sim \N(0, \sigma_{i,h}^2),
\label{eq:lp}
\end{equation}
for $t = p+1, \ldots, T_i - h$, providing $T_{i,h} = \max \{T_i - p - h, 0\}$ observations for the $h$-step-ahead regression of series $i$. The structural impulse response of series $i$ at horizon $h$ is $\rho_{i,h}$ while $\bm{\beta}_{i,h}$ denotes the vector of parameters associated with the controls. The errors in \eqref{eq:lp} feature an MA($h-1$) structure when $h \geq 1$. At the impact horizon the regressions do not overlap, so serial correlation of this kind is absent and the HAC bandwidths introduced in Section~\ref{sec:muller} are set to zero at $h = 0$. We do not estimate this directly but construct a robust estimator for the error variance or the long-run variance and, in turn, the variance of  $\rho_{i,h}$.

\subsection{Hierarchical pooling for unbalanced panels of local projections}
Traditional estimation of LPs revolves around estimating all series by OLS but with robust standard errors to correct for the MA structure in the errors and possible heteroskedasticity. If $T_i$ is small, however, inference becomes notoriously unreliable, and responses at longer horizons cannot be estimated at all once $h$ exceeds $H_i$. We address these issues by setting up a joint model that pools many long and short time series together and exploits the cross-sectional dimension to deliver reliable estimates of $\rho_{i,h}$ for the short series.

Before discussing our actual proposal, it is worthwhile to consider a simple pooling specification. For simplicity, we focus on the case without exogenous controls and assume that the panel is characterized by one short time series $i$ and many long time series (so that $T_i \ll T_j$ for all $j \neq i$). The LP regression for the short series $i$ is then given by:
\begin{equation*}
    y_{i, t+h} = \rho_{i,h} w_t + u_{i, t+h}.
\end{equation*}
Let $\bm Y_{i,h}$ and $\bm w_{i,h}$ denote the corresponding full data vectors with $T_{i,h}$ rows. If $T_{i,h}$ is small, coming up with reliable estimates is hard. Our solution is to pool information through a hierarchical Bayesian model. As a simple starting point, we assume that $\rho_{i,h}$ comes from a common distribution:
\begin{equation}
    \rho_{i,h} \sim \mathcal{N}(\mu_h, {\tau}_h^2), \label{eq: common}
\end{equation}
where  $\mu_h$ can be interpreted as the average response of the endogenous series in the panel to a unit increase in $w_t$ and ${\tau}_h^2$ is the prior variance. Equation~\eqref{eq: common} can be interpreted as a Bayesian prior. Assuming that the error variance is known (and fixing the hyperparameters) yields the following posterior mean:
\begin{equation*}
    \overline{\rho}_{i,h} = \overline{\tau}_{i,h}^2 \left( w_{\text{data}} \cdot \hat{\rho}_{i,h} + w_{\text{prior}}\cdot \mu_h \right),
\end{equation*}
where $\overline{\tau}_{i,h}^2 = (V_{i,h}^{-1} + 1/{\tau}^2_h)^{-1}$ denotes the posterior variance, $\hat{\rho}_{i,h} = (\bm w_{i,h}' \bm w_{i,h})^{-1} \bm w_{i,h}' \bm Y_{i,h}$  the OLS estimator of $\rho_{i,h}$, and $V_{i,h}=\sigma_{i,h}^2 (\bm w'_{i,h} \bm w_{i,h})^{-1}$ the OLS variance. The non-normalized weights are given by $w_{\text{data}} = \bm w_{i,h}' \bm w_{i,h}/\sigma_{i,h}^2 {= V_{i,h}^{-1}} = \sum_{t=p+1}^{T_i-h} w_t^2/\sigma_{i,h}^2$ and $w_{\text{prior}}= 1/{\tau}_h^2$, respectively. 
The data-based weights directly depend on $T_{i,h}$, so that for fixed ${\tau}_h^2$, we put little weight on the data-based estimate if $T_{i,h}$ is small. In the extreme case that $T_{i,h} = 0$ (the case where $h > H_i$), we have $\bm w_{i,h}'\bm w_{i,h} = 0$, the data weight vanishes, and the posterior mean reduces to the prior mean.

We use this well-known result that the posterior mean is a  combination of the data-based estimate ($\hat{\rho}_{i,h}$) and the prior mean ($\mu_h$) to improve our estimate of $\rho_{i,h}$, since $\mu_h$, if estimated, contains information from the other series that have more data. We assume that $\mu_h$ arises from:
\begin{equation*}
    \mu_h \sim \mathcal{N}(0, \xi^2),
\end{equation*}
with prior variance $\xi^2$. Combining this prior with the conditional likelihood in (\ref{eq: common}) yields the posterior:
\begin{equation*}
    \mu_h \mid \{\rho_{i,h}\}_{i=1}^M \sim \mathcal{N}(\overline{\mu}_h, \overline{\xi}^2),
\end{equation*}
where {$\overline{\xi}^2 = (M/\tau_h^2 + 1/\xi^2)^{-1}$ and $\overline{\mu}_h = \overline{\xi}^2 \sum_{i=1}^M \rho_{i,h}/\tau_h^2$; the latter} reduces to the arithmetic mean $\sum_{i=1}^M \rho_{i,h}/M$ as the prior on $\mu_h$ becomes uninformative ($\xi^2 \to \infty$). Intuitively, our series-specific estimates are informed by the average response of other series. Even if the prior variance is kept fixed, the weight on the prior increases if the number of observations becomes small. If $T_{i,h} = 0$, the posterior mean of $\rho_{i,h}$ equals $\mu_h$.

For this specification to work well, we need to assume that the dynamic properties of $y_{i,t+h}$ with respect to movements in $w_t$ have some similarities across $i$. This holds if, e.g., we are interested in modeling the dynamic reactions of a large panel of different price series to a monetary policy shock. However, in many cases, we are interested in understanding the responses of many different series. If $\rho_{i,h}$ differs strongly from $\rho_{j,h}$, then the pooling solution ends up shrinking towards the wrong location. To cope with this, our solution is to use a mixture model that allows for shrinkage towards mixture-specific responses. We will discuss this model next.

\subsection{Sparse finite mixtures for local projections}
If the dynamic properties of the series are heterogeneous but there are groups of series that are similar, one could search for similar series and group them together to estimate a group-specific location parameter for $\rho_{i,h}$. Let $z_i \in \{1, \ldots, S\}$ denote a group/cluster assignment that is common across horizons. $z_i$ groups $y_{i,t+h}$ into one of $S$ clusters. Within each cluster, we expect that the responses to a shock $w_t$ are similar. Formally, instead of using a single Gaussian distribution, we propose a mixture of Gaussians with $S$ components:
\begin{equation*}
    \rho_{i,h} \sim \sum_{s=1}^S \pi_s \mathcal{N}(\mu_{s, h}, \tau_{s, h}^2),
\end{equation*}
where $\mu_{s, h}$ and $\tau_{s, h}^2$ are cluster-specific means and variances, respectively. $\pi_s$ is a weight associated with component $s$ so that $\text{Prob}(z_i = s) = \pi_s$.

A more convenient way of writing up the mixture is to use $z_i$ directly:
\begin{equation}
    \rho_{i,h} \mid z_i = s \sim \mathcal{N}(\mu_{s, h}, \tau_{s, h}^2), \label{eq: mixture}
\end{equation}
which states that if series $i$ belongs to group $s$, its response is informed only by the other members of that group and is shrunk towards the group mean $\mu_{s,h}$. Two features of this prior deserve emphasis. First, the prior is \emph{Gaussian}. So $\rho_{i,h}$ is pulled towards a data-estimated cluster center $\mu_{s,h}$ and \emph{not} towards zero. The prior borrows strength across similar series. It is not a device for sparsity or variable selection.  Second, and relatedly, the Gaussian form keeps the model conditionally conjugate, so that the posterior mean of $\rho_{i,h}$ under the mixture retains the transparent weighted-average structure of the single-pool case:
\begin{equation*}
    \overline{\rho}_{i,h} = \overline{\tau}_{i,h}^2(z_i) \left(w_{\text{data}} \hat{\rho}_{i,h} + w_{\text{prior}}(z_i) \mu_{s, h}\right).
\end{equation*}
Here, $\overline{\tau}_{i,h}^2(z_i) = (\bm w_{i,h}' \bm w_{i,h}/\sigma^2_{i,h} + 1/\tau^2_{z_i,h})^{-1}$ is the posterior variance of series $i$ which now depends on the cluster allocation $z_i$.  The prior-based weight is now given by $1/\tau^2_{z_i,h}$ and also depends on the cluster membership. Notice that, for a fixed component-specific prior variance $\tau^2_{z_i, h}$, the weight on the prior rises sharply if $T_{i,h}$ becomes small. In the limiting case of $T_{i,h} = 0$ we end up imputing the dynamic response of $y_{i,t}$ to a shock $w_t$ using the component-specific prior mean (as opposed to a common prior mean in the pooling case). This allows us to borrow information from other time series in a (possibly) huge macroeconomic time series panel.

Series that are, at first glance, only loosely related can still be similar in terms of their full IRF profile. The model exploits this by grouping series that have similar responses into a single cluster. If the number of series per cluster, $M_s = \#\{i : z_i = s\}$, is small and $T_{i,h}$ is small as well then this would imply that only few series inform the IRF estimates. In the case where $M_s$ is large but $T_{i,h}$ is small we end up with posterior estimates for $\rho_{i,h}$ that are closer to the group-specific mean and that are estimated {with considerable precision, since the cross-sectional information is strong}.

The case of small $M_s$ and $T_{i,h}$ deserves further attention. If the number of series within a group is small (or, in the extreme case, $M_s = 1$), it raises the question of whether our approach still produces efficiency gains, given that there is then little (or no) cross-sectional information within the cluster to exploit. Taken literally, a singleton cluster would leave $\rho_{i,h}$ informed only by the $T_{i,h}$ observations of series $i$ itself, shrunk towards a cluster mean $\mu_{s,h}$ that is in turn determined by that very series, which is precisely the situation the pooling was meant to improve upon.

The third level of the hierarchy addresses this. Rather than treating the cluster means as free parameters, we let them arise from a common population distribution,
\begin{align}
\mu_{s,h}  \mid m_h, B_h^2 &\sim \mathcal{N}(m_h, B_h^2), \label{eq:level2}\\
m_h &\sim \mathcal{N}(0, c \cdot s_y^2), \label{eq:level3}
\end{align}
where $m_h$ is a population center shared by all clusters and $B_h^2$ governs how far an individual cluster is allowed to depart from it.  {In the prior on $m_h$, $s_y^2$ denotes the sample variance of the outcome series (equal to one, since every series is standardized before estimation) and $c$ is a large constant, set to $c = 100$ throughout, so that the prior on the population center is weakly informative.} Both prior variances, $\tau_{s,h}^2$ and $B_h^2$, feature weakly informative inverse Gamma priors.  A cluster that includes many similar series pulls $\mu_{s,h}$ towards its own within-group mean and is only mildly  influenced by $m_h$.  A small or singleton cluster has little of its own to contribute and is instead pulled back towards $m_h$, which aggregates information across all remaining series. The borrowing of strength thus never switches off entirely: what a series cannot learn from its own group, it still learns  from the full panel.

Two implications follow. First, the efficiency gain is not automatic. It depends on the series we aim to pool and their time series properties. The gains are largest when series cluster tightly, whereas if a series is unlike every other the mixture is free to isolate it in its own component, so that no spurious \emph{within-cluster} precision is imposed on it.  But an isolated series still borrows from the panel. Its cluster mean is pulled towards $m_h$ by \eqref{eq:level2}, so information is obtained through the population center. Only when $B_h^2$ is large or the series is long does its estimate become purely series-specific. What isolation buys is protection against being pulled towards a group the series does not belong to. Second, the extra layer also sharpens cluster identification. Because the cluster means are anchored to a common center, the sampler is less prone to producing near-empty components that accommodate a handful of outliers{, and, if the data support it, the model collapses cleanly to the single-cluster pool of the previous subsection}.

\subsection{Pooling the coefficients on the controls}\label{sec:jointpool}
The framework so far pools only the object of primary interest, the response coefficient $\rho_{i,h}$, and leaves the control coefficients $\bm\beta_{i,h}$ to be estimated series-by-series. This is a sensible default, since $\rho_{i,h}$ is the coefficient the researcher usually cares about. But the very argument that makes pooling helpful for $\rho_{i,h}$ applies just as well to the controls since these are estimated on the same short samples, and series that respond similarly to the shock are also likely to share similar control dynamics. Because $\bm\beta_{i,h}$ can be high-dimensional, the potential gain from pooling  can be larger than for the single response coefficient.  That is especially the case if the researcher follows common advice (e.g., \citealp{Oleaetal2025}) and opts for a generous set of controls in the interest of robustness.

We therefore also offer a richer variant that uses a mixture for all coefficients in the model. Collecting the response and the controls in $\bm\theta_{i,h} = (\rho_{i,h}, \bm\beta_{i,h}')'$ of dimension $d = 1 + k$, the within-cluster prior \eqref{eq: mixture} generalizes to
\begin{equation}
\bm\theta_{i,h} \mid z_i = s \sim \mathcal{N}(\bm\mu_{s,h}, \tau_{s,h}^2 \mathbf{I}_d), \label{eq:jointpool}
\end{equation}
so that every series in cluster $s$ is now shrunk towards a common cluster mean $\bm\mu_{s,h}$, which is now a vector consisting of a response part and a control part, as opposed to a scalar. The within-cluster scaling parameter $\tau_{s,h}^2$ is shared across coefficients, so that the dynamic responses and each control coefficient are pulled towards the cluster mean with equal strength. This is innocuous because we standardize all regressors before estimation so that equal shrinkage on the standardized scale amounts to covariate-specific shrinkage in the original units, scaled by each covariate's own standard deviation.

We assume that the cluster-specific means arise from a common distribution. To allow for the possibility that some coefficients are homogeneous across clusters, we use a shrinkage prior to pull cluster means towards a common location. Formally, {writing $\mu_{s,h,j}$ for the entry of $\bm\mu_{s,h}$ associated with control $j$,} we use a horseshoe prior:
\begin{equation}
\mu_{s,h,j} \mid m_{h,j}, \psi_j, \psi_B \sim \mathcal{N}\bigl(m_{h,j}, \psi_j^2 \psi_B^2\bigr), \qquad
\psi_j \sim \mathcal{C}^+(0,1), \quad \psi_B \sim \mathcal{C}^+(0,1), \label{eq:level2-beta}
\end{equation}
for $j = 1, \ldots, k$, with a mildly informative prior $m_{h,j} \sim \mathcal{N}(0, c \cdot s_y^2)$ on the center. Here $\psi_B$ is a global scale that induces shrinkage across the controls $j = 1, \ldots, k$ and $\psi_j$ a control-specific local scale, shared across clusters and horizons. We deliberately leave the response coefficient $\rho_{i,h}$ out of the horseshoe hierarchy. The spread of the cluster means in the response is still governed by $B_h^2$ in \eqref{eq:level2}. The horseshoe shrinks aggressively, which is suitable for nuisance coefficients but not the object we care about (i.e. the IRFs).  Coefficients on which the clusters agree have their cluster means pulled towards a common location, recovering complete pooling for those controls whose coefficients are homogeneous across series.

\subsection{The prior on the weights and the number of clusters}
So far we have remained silent  about how many clusters $S$ to use, or how the weights $\pi_s = \mathrm{Prob}(z_i = s)$ are chosen. We do not wish to fix the number of active groups in advance, since it is rarely known and is arguably the object we would  like to learn from the data. Therefore, we follow the sparse finite mixture approach of \citet{malsinerwalli2016} and set $S$ to a deliberately generous upper bound (in our applications, we fix $S = 8$) while setting up a shrinkage prior that empties out irrelevant components, effectively leading to a specification with fewer than $S$ clusters.

We model allocations and weights  as:
\begin{align*}
z_i \mid \bm{\pi} &\sim \mathrm{Cat}(\bm{\pi}),\\
\bm{\pi} \mid e_0 &\sim \mathrm{Dir}(e_0, \ldots, e_0),\\
e_0 &\sim \mathrm{Gamma}(a_e, b_e),
\end{align*}
with a symmetric Dirichlet prior on the weights whose concentration parameter $e_0$ is itself estimated.

The mechanism, developed by \citet{malsinerwalli2016} based on the overfitting-mixture asymptotics of \citet{rousseau2011asymptotic}, is as follows. When $e_0$ is small, the Dirichlet prior places its mass near the vertices of the simplex and so strongly favors weight vectors in which most entries are close to zero: irrelevant components are emptied out, while the components the data genuinely require retain appreciable weight. The effective number of groups, $S^* = \#\{s : M_s > 0\}$, is thus obtained as a by-product rather than set as a tuning choice, and placing a Gamma hyperprior on $e_0$ lets the data calibrate how aggressively the pruning operates.

This construction keeps the dimension of the parameter space fixed, so that estimation proceeds through a standard Gibbs sampler rather than a transdimensional scheme that has to add and delete components. It also nests the simple Gaussian pool with which we began as the special case in which the prior empties all but one component, so the model chooses between one common response and several group-specific responses endogenously rather than by assumption.

Estimation is carried out using a Markov chain Monte Carlo (MCMC) sampler that exploits the fact that most conditional distributions take a well known form and hence are amenable to Gibbs sampling.  We provide further details on the full conditionals and the complete algorithm in Appendix~\ref{app:fullcond}. 

The mixture model is not identified due to label switching. To solve this, we use the random permutation sampler of \cite{fruhwirth2001markov} to ensure that the sampler explores the full posterior distribution. After MCMC sampling, we identify cluster-specific quantities such as $\bm \mu_{s,h}$, cluster sizes $M_s$ or individual allocations $z_i$ using the $K$-means-based relabeling procedure detailed in   \citet{malsinerwalli2016}.

\subsection{Robustification of the error variances}\label{sec:muller}
The LP in \eqref{eq:lp} is not a generative model since $u_{i,t+h}$ is a projection error. Moreover, it is heteroskedastic and, under a VAR-type DGP, serially correlated with an MA($h-1$) structure. Neither biases the point estimates, but both make the posterior variance too small. Following \citet{muller2013}, we rescale the posterior draws of $\rho_{i,h}$, substituting a HAC long-run variance of the score for the posterior one, so that the resulting credible sets are calibrated to a sandwich variance. This makes them approximately valid in the frequentist sense.

Unit-by-unit, this variance is far too noisy in our samples. We therefore pool it over the partition $\{z_i\}$ the mixture delivers. When units in a cluster share an IRF profile (and the DGP is VAR-like), the LP residual's MA coefficients are functions of that profile, so units that respond alike might have similar long-run variances. With $g_{i,t,h} = \tilde{w}_{i,t} \hat{u}_{i,t+h}$ denoting the LP score{, where $\tilde{w}_{i,t}$ is the shock residualized on the controls $\bm x_t$ of series $i$,} $\hat{\Gamma}_{i,h}(\ell) = T_{i,h}^{-1}\sum_t g_{i,t,h} g_{i,t-\ell,h}$ its autocovariance at lag $\ell$, and $\mathcal{I}_{s,h} = \{i : z_i = s, H_i \ge h\}$, the unit-specific and pooled HAC variances are given by:
\begin{equation}
\hat{J}_{i,h} = \hat{\Gamma}_{i,h}(0) + \sum_{\ell=1}^{L_{i,h}} \Bigl(1 - \frac{\ell}{L_{i,h} + 1}\Bigr) \cdot 2 \hat{\Gamma}_{i,h}(\ell), \qquad
\hat{J}_{s,h} = \frac{\sum_{i \in \mathcal{I}_{s,h}} T_{i,h} \hat{J}_{i,h}}{\sum_{i \in \mathcal{I}_{s,h}} T_{i,h}}.
\label{eq:NW_pooled}
\end{equation}
The bandwidth $L_{i,h} = \min\{{\max\{h-1, 0\}},\ T_{i,h}-1\}$ is implied by the horizon-$h$ window overlap rather than tuned; the truncation binds only for series too short for $h-1$ lags. The outer $\max$ sets $L_{i,0} = 0$ at the impact horizon. There the sum in \eqref{eq:NW_pooled} is empty and $\hat{J}_{i,0}$ is just the heteroskedasticity-robust $\hat{\Gamma}_{i,0}(0)$. The $T_{i,h}$-weighted average treats the cluster's scores as one sample. Short units inherit an estimate dominated by better-measured neighbors while long units dominate their own. We then rescale the draws to match the sandwich variance from $\hat{J}_{s,h}$ rather than the posterior variance $\overline{V}_{\mathrm{post}, i, h}$ that the sampler itself delivers,
\begin{equation*}
\rho_{i,h}^{(r), \mathrm{adj}} = \overline{\rho}_{i,h} + \kappa_{i,h} \bigl(\rho_{i,h}^{(r)} - \overline{\rho}_{i,h}\bigr), \quad
\kappa_{i,h} = \sqrt{\frac{\hat{V}_{\mathrm{sand}, i,h}^{\mathrm{pool}}}{\overline{V}_{\mathrm{post}, i, h}}}, \quad
\hat{V}_{\mathrm{sand}, i,h}^{\mathrm{pool}} = \frac{T_{i,h} \hat{J}_{s,h}}{\bigl(\tilde{\bm{w}}_{i,h}'\tilde{\bm{w}}_{i,h}\bigr)^{2}},
\end{equation*}
where $r = 1, \ldots, R$ indexes the retained draws, $\tilde{\bm{w}}_{i,h}$ stacks the residualized shock, and
\begin{equation*}
\overline{\rho}_{i,h} = \frac{1}{R} \sum_{r=1}^{R} \rho_{i,h}^{(r)}, \qquad
\overline{V}_{\mathrm{post}, i, h} = \frac{1}{R} \sum_{r=1}^{R} \bigl(\rho_{i,h}^{(r)} - \overline{\rho}_{i,h}\bigr)^{2}
\end{equation*}
are the posterior mean and variance of $\rho_{i,h}$ implied by the sampler.\footnote{Both are computed from the retained draws and are therefore marginal over the allocation $z_i$ and the hierarchical parameters. It is this marginal spread that the rescaling replaces, so that the adjusted draws carry sample variance exactly $\hat{V}_{\mathrm{sand}, i,h}^{\mathrm{pool}}$.} This correction does not impact the point estimates but only affects their uncertainty estimates.

One case needs an explicit convention. If $h > H_i$ then $T_{i,h} = 0$, the residualized shock vector is empty, and $\hat{V}_{\mathrm{sand}, i,h}^{\mathrm{pool}}$ is a ratio of two zeros. The sandwich is undefined. We then set $\kappa_{i,h} = 1$ and leave the draws unadjusted{, and we do the same in the rare case where a series has data but its residualized shock is nearly collinear with the controls. This is coherent: at such horizons the posterior is a prediction from the hierarchy rather than a rescaled sampling distribution.} Our reported intervals are therefore of two kinds. Where a series has data they are sandwich-calibrated, and beyond that they are purely hierarchical.

The same device gives us a band for the cluster average reported in Section~\ref{sec:app-cluster-irfs}.  Its OLS counterpart is the average of the unit-specific estimators over the $M_{s,h} = |\mathcal{I}_{s,h}|$ members with data at horizon $h$. Since each unit's estimation error is the sum over $t$ of the scaled scores $g_{i,t,h}/(\tilde{\bm{w}}_{i,h}'\tilde{\bm{w}}_{i,h})$, the score of the average at date $t$ is $g_{s,t,h} = M_{s,h}^{-1} \sum_{i \in \mathcal{I}_{s,h}} g_{i,t,h} / (\tilde{\bm{w}}_{i,h}'\tilde{\bm{w}}_{i,h})$. We apply \eqref{eq:NW_pooled} to $g_{s,t,h}$ in place of $g_{i,t,h}$, which returns a cluster long-run variance $\hat{J}^{A}_{s,h}$ and a multiplier $\kappa_{s,h}$. Members with no data at horizon $h$ are handled by the convention of the previous paragraph. Adding up the members' scores date by date, before taking the lag sum, is what carries the covariance between series in the same cluster into the cluster band.\footnote{This is the panel long-run variance estimator of \citet{driscoll1998}. Aggregating first is what makes it practical here. It replaces $M_s(M_s+1)/2$ pairwise long-run covariances with a single Newey--West sum.}

\section{Simulation evidence}\label{sec:simulation}
This section presents simulation evidence based on a realistic DGP calibrated to FRED-MD monthly data. We evaluate the variants of our LP estimator against a textbook OLS LP on point estimation accuracy, estimation uncertainty and interval coverage.  We first discuss the design of our DGP before considering the results of our simulation exercise.

\subsection{A data-generating process calibrated to monthly US macro data}\label{sec:sim-dgp}
We consider a realistic DGP that resembles dynamics observed in US macroeconomic datasets. To this end, we use the FRED-MD dataset of  \citet{mccracken2016fred} and use the 2024-06 vintage but focus on the period from 1965M1--2019M12 ($660$ months). 

We assume that the time series in the panel are well described by four factors. These are extracted as the first principal component from the FRED-MD categories real activity (output and income), employment (labor market), prices, and housing. Before factor extraction, we standardize the time series from which we extract the factors and make sure that series with more than $5\%$ missing observations are being excluded.   {The shock of interest $w_t$ is a monetary policy shock. To calibrate it, we estimate a seven-variable VAR($12$) on the four group factors, the federal funds rate, and two additional factors extracted from the FRED-MD money/credit and stock-market categories. The shock is identified recursively, with the funds rate ordered after the four slow-moving group factors and before the money/credit and stock-market factors.}

Each factor follows an autoregressive distributed lag model:
\begin{equation}
F_{g,t} = a_{g,1} F_{g,t-1} + a_{g,2} F_{g,t-2}
        + \sum_{\ell=0}^{2} b_{g,\ell} w_{t-\ell}
        + \sigma_g \eta_{g,t},\qquad
\eta_{g,t}\stackrel{\mathrm{iid}}{\sim}\N(0,1),
\label{eq:sim-factor}
\end{equation}
whose coefficients $(\bm a^{(g)}, \bm b^{(g)}, \sigma_g)$ are based on the OLS estimates of the empirical factor $F^{\text{emp}}_{g,t}$ on its two own lags and the current and two lagged values of $w_t$. Three adjustments improve the calibration of the DGP with respect to matching common data properties in LP analysis and having meaningful impulse response functions in the DGP.  {First, for real activity and employment, we increase the estimated first-order coefficient $a_{g,1}$ by five percent, since the fitted low-order specification otherwise produces factor paths that are less persistent than their empirical counterparts.} Second, because the fitted innovation scale $\sigma_g$ also absorbs measurement error and shocks we do not model, we halve it, so that the component of each factor driven by $w_t$ is not corrupted by idiosyncratic factor noise. Finally,  we scale the calibrated shock loadings $b_{g,\ell}$ up by a factor of three. This choice ensures that the monetary policy shock has important quantitative effects on the factors. 

These factors drive the individual time series. This is achieved by generating $M=80$ observed series as a linear function of the factors.  We assume four clusters of $20$ series each, where the factor loadings are given by:
\begin{equation*}
\lambda_{i,z_i} = 1 + \xi_{i,z_i},\quad
\lambda_{i,g} = \xi_{i,g}\ (g \ne z_i),\quad
\xi_{i,\cdot}\stackrel{\mathrm{iid}}{\sim}\N(0,\sigma_\lambda^2),\quad \sigma_\lambda = 0.05,
\end{equation*}
so that $y_{i,t} = \bm\lambda_i'\bm F_t + \sigma_e\varepsilon_{i,t}$ with a small idiosyncratic noise $\sigma_e = 0.05 \cdot \mathrm{median}(\sigma_g)$ and $\varepsilon_{i,t}\stackrel{\mathrm{iid}}{\sim}\N(0,1)$. This design implies that each series predominantly loads on the factor within that group but it also has loadings close to zero for the other factors outside a particular group.  

To analyze how our algorithm performs across time series of different lengths, we set up the panel to be unbalanced. Half of the $M=80$ units are \emph{long}, with $T_i = 500$ months, and the other half are much shorter. We run the design twice. In the \emph{short} design the lengths of the shorter units are drawn uniformly from $\{100,\ldots,150\}$, in the \emph{very short} design uniformly from $\{25,\ldots,60\}$. The two designs are otherwise identical. We use the same loadings $\bm\lambda_i$, the same long units, the same short/long assignment, and the same shocks. The only difference arises from the short units' sample lengths, which lets us trace performance as the short series become progressively less informative. Results for the long units are reported from the short design. 
We initialize the DGP as follows. Each factor recursion in \eqref{eq:sim-factor} is started at zero and we simulate $550$ observations, with the first $50$ observations then dropped to avoid that the initialization impacts the time series dynamics. The shock is drawn from $w_t \stackrel{\mathrm{iid}}{\sim} \N(0,1)$. The empirical  estimate of $w_t$ above is used only to calibrate the loadings $b_{g,\ell}$.  The loadings $\bm\lambda_i$, the long/short assignment, and the sample lengths $\{T_i\}$ are drawn once and then held fixed across the $n_{\mathrm{MC}}=500$ Monte Carlo replications from the DGP.  Only the innovations $\{w_t\}$, $\{\eta_{g,t}\}$, and $\{\varepsilon_{i,t}\}$ differ across replications. This isolates the sampling variability of the estimators from variability in the panel design itself.

\subsection{Competing models and prior setup}\label{sec:sim-estimators}
We compare five estimators of the unit-level response $\rho_{i,h}$ that can be viewed as nested alternatives of our proposed model.  The benchmark is \texttt{Naive LP}, a per-unit OLS LP with Newey--West (HAC) standard errors at the unit level. The HAC bandwidth is tied to the horizon to match the MA($h-1$) structure of the LP residual, $L_{i,h} = \min\{\max\{h-1, 0\},\ T_{i,h}-1\}$, with the Bartlett weights of \eqref{eq:NW_pooled}; the truncation binds only for the very short series at long horizons. This is the estimator an applied researcher would run series-by-series, and the one the pooling is meant to improve upon.\footnote{We also consider Huber--White  standard errors, which remain asymptotically valid here because every regression is lag-augmented \citep{montielolea2021}. In this case, coverage rates are very close to the HAC coverage rates.}

We consider four Bayesian estimators.  The first is our general framework that uses a mixture pool for  $\rho$ and $\bm \beta$, labeled as \texttt{Pool: $(\rho,\bm\beta)'$ (SFM)} in the subsequent figures and tables.  The other three can be considered special cases of the general estimator. We let \texttt{Pool: $\rho$ ($S{=}1$)} denote the single-component Gaussian pool of Section~\ref{sec:framework}, which shrinks each series' response towards one common, panel-wide mean to isolate the merits of cross-sectional pooling before any clustering. \texttt{Pool: $\rho$ (SFM)} replaces this single mean specification with the sparse finite mixture (with an upper bound of $S = 8$ components), so that responses are shrunk towards \emph{cluster}-specific means. The comparison with the previous estimator isolates the contribution of letting the pooling target be group-specific rather than global. The last  estimator \texttt{Pool: $(\rho,\bm\beta)'$ ($S{=}1$)} additionally pools the control coefficients  towards a single global mean, disentangling the gain from pooling $\bm\beta$ from the gain from cluster-specific pooling of $\rho$.

All four Bayesian estimators are sampled using the Gibbs algorithm detailed in the Appendix. We retain $5{,}000$ draws after discarding the first $5{,}000$ draws as burn-in.  Cluster allocations are initialized by $K$-means on the standardized $y$-series (each unit's last $\min_i T_i$ observations), and  label switching is dealt with using the approach put forth in \cite{malsinerwalli2016}.  

The hierarchical priors follow the setup of \citet{malsinerwalli2016} and are kept mildly informative and data-scaled, with the variance scales recalibrated in every replication from that replication's own data. {Both variance scales come from a preliminary four-group $K$-means partition of the standardized series, computed on the long units alone. For each preliminary group with at least two long units and each horizon, we compute the sample variance of the OLS estimates $\hat\rho_{i,h}$ across the units in that group; $\hat{v}_{\text{within}}$ is the median of these variances over all groups and horizons, and $\hat{v}_{\text{between}}$ is defined analogously using the variance of the group means. The within-cluster variance then has prior $\tau_{s,h}^2 \sim \mathcal{IG}(a_0, b_0)$ with $a_0 = 2.5$ and scale $b_0 = (a_0 - 1) \hat{v}_{\text{within}}$, and the between-cluster variance has prior $B_h^2 \sim \mathcal{IG}(a_B, b_B)$ with $a_B = 2.5$ and scale $b_B = (a_B - 1) \hat{v}_{\text{between}}$, so that each prior mean equals the corresponding dispersion estimate.} The mixture weights carry a sparse Dirichlet prior whose concentration $e_0$ is not fixed but estimated by the Metropolis step of Appendix~\ref{app:fullcond}, under a $\mathrm{Gamma}(1, 200)$ prior with mean $0.005$, so the data decides how aggressively redundant components are emptied out.

\subsection{Point estimation accuracy across horizons}\label{sec:sim-rmse}
Table \ref{tab:mae-headline} reports mean absolute errors (MAEs) against the true $\rho_{i,h}$ on data-informed unit-horizon pairs ($h \le H_i$), expressed relative to the naive LP benchmark. The shaded naive LP row reports its raw MAE while all other rows are ratios so that values smaller than one imply more accurate point estimation than the benchmark. We report them separately for three series-length regimes: very short ($T \in [25, 60]$), short ($T \in [100, 150]$), and long ($T = 500$). Results are further disaggregated into three horizon buckets, for which we report averages: \emph{short-run} ($h = 0,\ldots,4$), \emph{medium-run} ($h = 5,\ldots,12$), and \emph{long-run} ($h = 13,\ldots,24$). In the short and long regimes every data-informed pair is also one where naive LP returns a value, so the comparison is apples-to-apples; in the very-short regime naive LP's design exceeds the available sample for most unit-horizon pairs.\footnote{\label{fn:naive-veryshort}The unit-specific design has $29$ coefficients ($p = 12$ lags of $y$ and $w$, intercept, shock, three factor controls), so with $T \in [25, 60]$ naive LP is computable on only $19\%$ of the data-informed unit-horizon pairs ($46\%$/$23\%$/$2\%$ across the three horizon buckets).  Its very-short entries in Table~\ref{tab:mae-headline} are the MAE over those pairs (and hence the denominators of the very-short ratios), while the pooled entries cover all data-informed pairs.}

\input{tables/tab_mae_headline.tex}

Starting with the short time series case (results in the middle columns of Table~\ref{tab:mae-headline}), the approach of pooling estimates of the LP coefficients but not the controls significantly reduces the estimation error of estimates relative to naive LP estimates.  With a single cluster, the  reduction rises from about 30 percent at short horizons to nearly 50 percent at longer horizons.  Allowing a larger number of clusters further improves LP accuracy at short and medium horizons, but not long horizons.  Pooling both the LP and controls' coefficients yields a slightly more pronounced reduction in MAEs, of roughly 50 percent at all horizons (with the larger number of clusters).  From this  perspective, with short time series, our baseline approach that pools for all coefficients and allows a larger number of clusters performs best among the pooling specifications.  While the single cluster estimators can rival the gains at long horizons, their sharp homogeneity restrictions come with costs at short and medium horizons.

In Table~\ref{tab:mae-headline}'s results, the relative impacts of pooling time series for other sample sizes align with what might be expected.  With very short time series, pooling has even larger benefits for point estimation accuracy, with our baseline specification reducing MAEs by about 80 percent at shorter and medium horizons and 90 percent at longer horizons.  With long time series, relative to a simple LP estimator that is relatively precise in large samples, pooling of estimates typically leads to some added estimation error.  With our baseline approach that pools for all coefficients and allows a larger number of clusters, this cost is fairly small (especially when gauged against the benefits achieved for smaller sample sizes), at roughly 6 percent.  We say estimation error rather than bias for a reason. Mean absolute error mixes systematic error with sampling dispersion, and Table~\ref{tab:mae-headline} cannot tell the two apart. The left panel of Figure~\ref{fig:bias-sd} shows that pooling does add bias on long series, and this is the likely source of the gap.  The cost can be more substantial with some of the other pooling estimators, most sharply for the estimator that pools all coefficients with the number of clusters limited to one, which results in an increase in MAEs between roughly 20 and 60 percent, depending on the horizon.

\subsection{The bias-variance relationship}\label{sec:sim-biasvar}

\begin{figure}[!tbp]\centering
\includegraphics[width=0.95\linewidth]{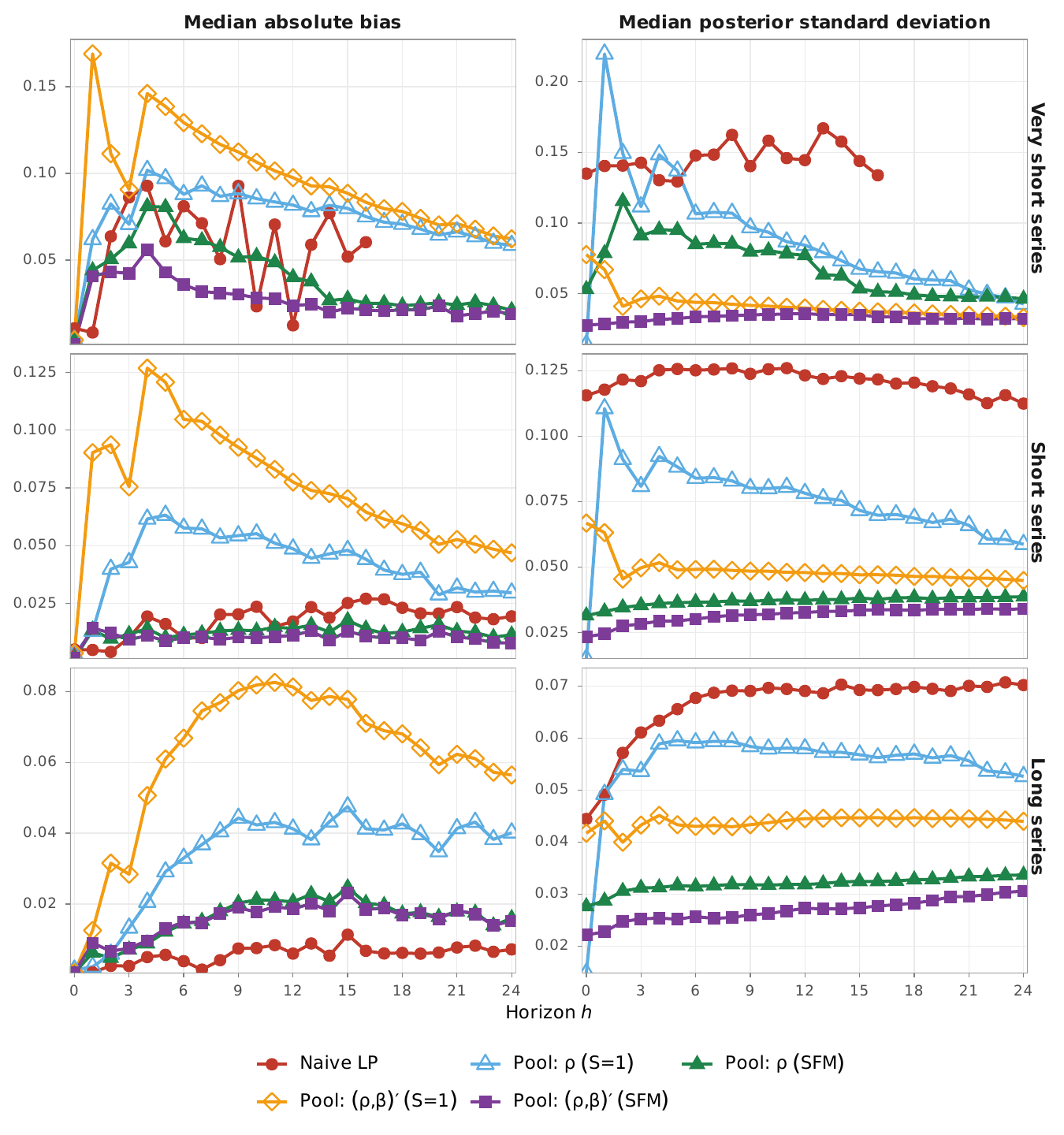}
\caption{Bias and posterior standard deviation by horizon, FRED-MD calibration.}
\label{fig:bias-sd}
\fignotes{Rows, top to bottom: very short ($T \in [25, 60]$), short ($T \in [100, 150]$), and long ($T = 500$) series; $y$-axis free per panel.  Left column: across-unit median of $\lvert \mathbb{E}_r[\hat\rho_{i,h}^{(r)}] - \rho_{i,h}\rvert$.  Right column: across-unit median of the posterior standard deviation of $\rho_{i,h}$, taking the median across replications per unit-horizon pair; for naive LP, its Newey--West standard error.  Both on data-informed unit-horizon pairs, naive LP drawn where it is computable (footnote~\ref{fn:naive-veryshort}).}
\end{figure}

Of course, in general, reducing bias can come at a cost of increasing variance.  To assess impacts on both bias and variance for our pooling approaches as compared to simple LP estimation, Figure~\ref{fig:bias-sd} reports the median absolute bias (left column) and the median posterior standard deviation (right column) by horizon, separately for very short, short, and long series and averaged over the replications from the DGP.  Both are computed on data-informed unit-horizon pairs; naive LP appears wherever it is computable.  The bias results in the left column of the figure align with the evidence in Table~\ref{tab:mae-headline}.  The standard deviation results in the right column shed new light on the alternative approaches.  Again starting with the short time series case, all of the pooling approaches reduce estimator variance compared to the simple LP estimator.  The lowest variance is achieved with the baseline approach that pools for all coefficients and allows a larger number of clusters, with the alternative of pooling just the LP coefficient nearly as precise.  With very short or long time series, this baseline approach continues to yield variances lower than or as low as those of any other estimator, and it yields sharp reductions in variance compared to simple LP estimates.  Combined, the bias and standard deviation results also highlight that, among the pooling specifications, those with homogeneity restrictions are generally (if not quite uniformly) dominated by those that allow multiple clusters whose number is determined by the data.

To put the bias-variance tradeoff in perspective, following the approach of \citet{li2024}, we define the unit-specific loss
\begin{equation*}
L_{i,h}^m(\omega) = \omega \bigl(\widehat{\mathrm{bias}}_{i,h}^m\bigr)^2 + (1-\omega) \widehat{\mathrm{var}}_{i,h}^m,
\qquad \omega \in [0.5, 1.0],
\end{equation*}
where $\omega$ ranges from equal weighting ($\omega = 0.5$) to pure squared bias ($\omega = 1$).  As in Figure~\ref{fig:bias-sd}, we take $\widehat{\mathrm{var}}_{i,h}^m$ to be the posterior variance of $\rho_{i,h}$ (the median across replications; for naive LP, its squared Newey--West standard error), so the trade-off weighs bias against the uncertainty each method actually reports.

\begin{figure}[!tbp]\centering
\includegraphics[width=0.95\linewidth]{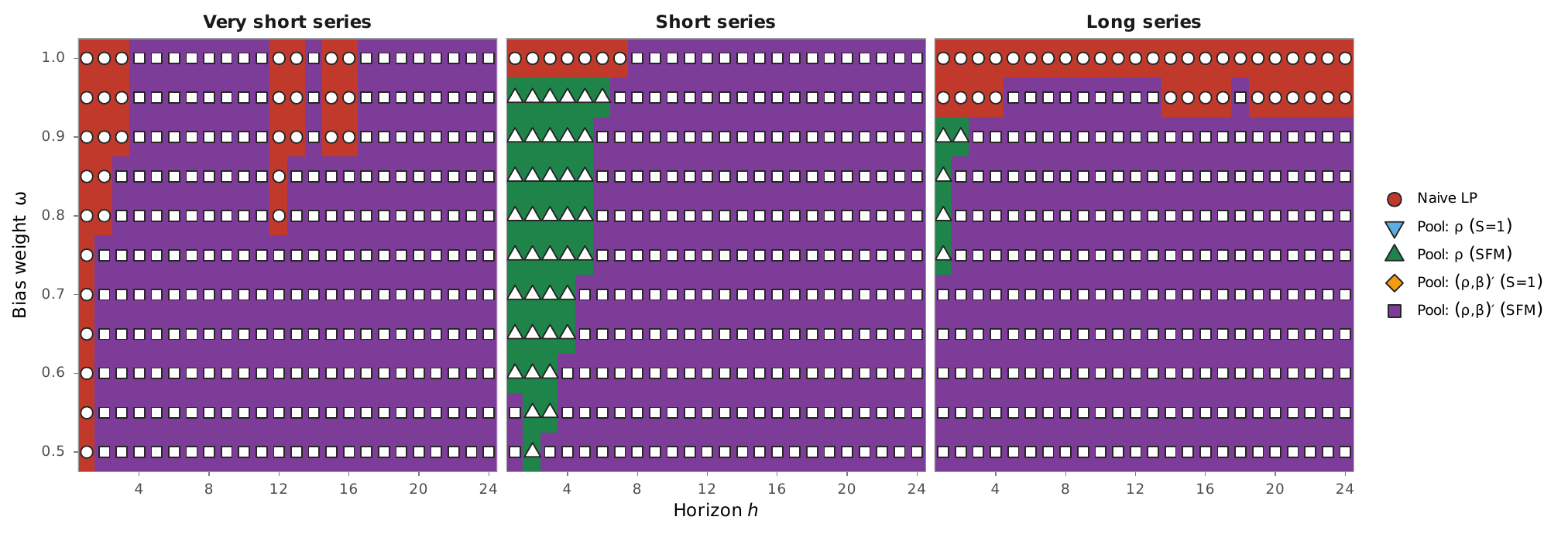}
\caption{Loss-minimizing estimator at each $(h, \omega)$ pair, FRED-MD calibration.}
\label{fig:bestmethod}
\fignotes{Panels, left to right: very short, short, and long series.  Color and per-cell symbol: the method minimizing the across-unit average loss $L_{i,h}^m(\omega)$ at each $(h, \omega)$ pair (the symbol identifies the method in grayscale).  Each method's loss is averaged over its data-informed unit-horizon pairs; naive LP competes where it is computable (footnote~\ref{fn:naive-veryshort}).}
\end{figure}

\begin{figure}[h!]\centering
\includegraphics[width=\linewidth]{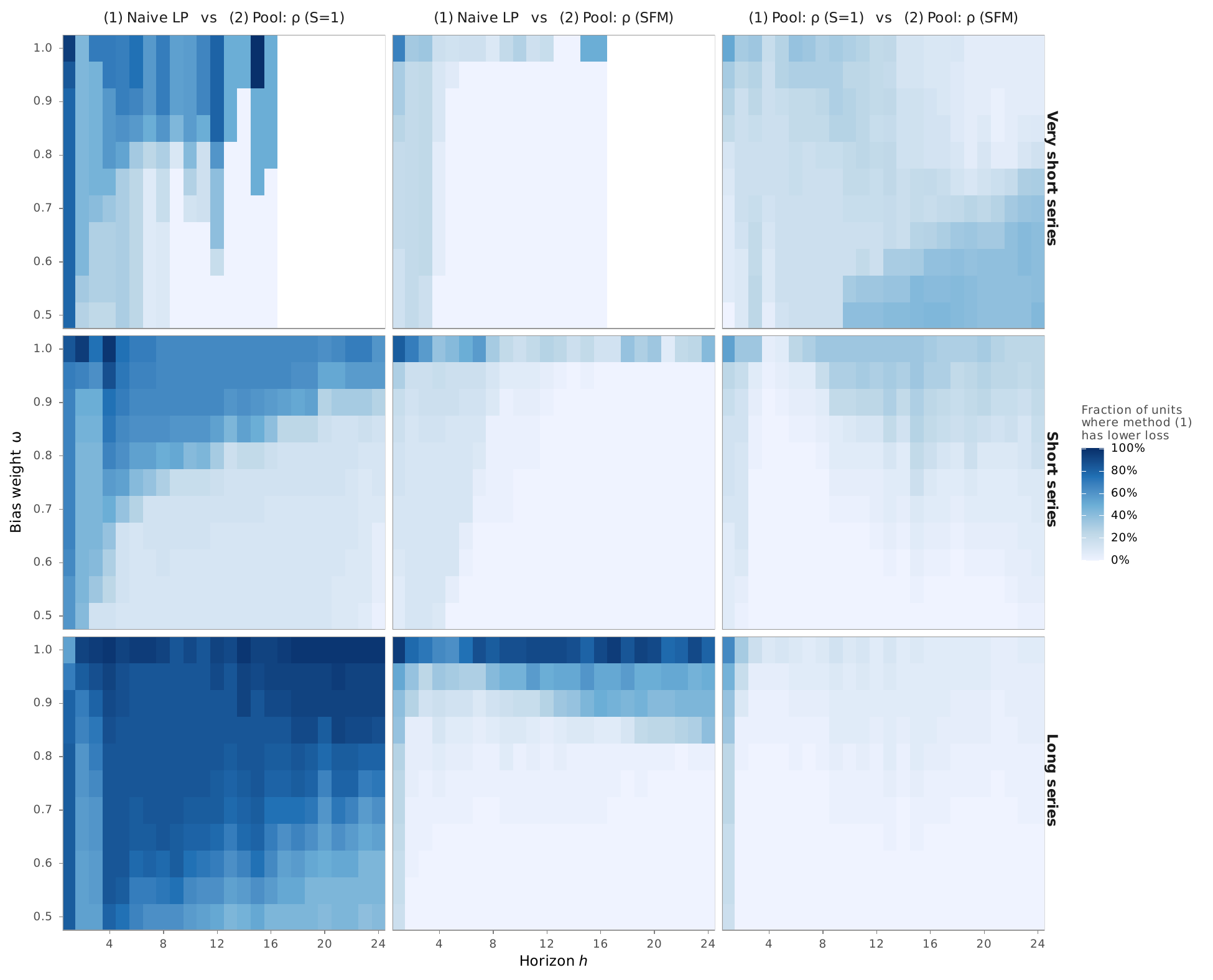}
\caption{Pairwise loss comparison, FRED-MD calibration.}
\label{fig:b4}
\fignotes{Each entry: the fraction of units at which method~(1) achieves a lower $L_{i,h}^m(\omega)$ than method~(2), on unit-horizon pairs where both are computable.  Rows, top to bottom: very short, short, and long series; columns: the three pairwise comparisons.  Darker blue = method (1) wins more often; blank regions in the very-short row are horizons at which naive LP is computable for no pair.}
\end{figure}

Figure~\ref{fig:bestmethod} reports the loss-minimizing estimator as a function of the horizon and bias weight, with separate panels for the three different sample sizes.  For all sample sizes, our baseline joint pooling approach broadly dominates across horizons and bias weights.  In the exceptions, the $\rho$-only SFM performs best at the shortest horizons with short series and a moderate-$\omega$ setting with long time series.  Simple LP estimation is preferable only with near-unity weights on bias; with some weight on variance, even with long time series, the bias advantage of simple LP is overcome by its disadvantage in estimator variance.  

To shed additional light on the relative loss performance of the alternative estimators, Figure \ref{fig:b4} provides pairwise comparisons, reporting percentages of cases in which one estimator yields lower loss than another, across horizons and loss weights.  The included comparisons are selected with an eye to illustrating performance of different pooling methods, given that  Figure~\ref{fig:bestmethod} has generally established the dominance of our baseline pooling estimator. (Note that simple  LP is computable nowhere beyond $h \approx 16$ in the very-short time series group, so the naive-LP columns of Figure~\ref{fig:b4} are blank there.) Starting with the simple LP versus the estimator that pools LP coefficients (not controls) and allows multiple clusters (results in the middle column), this pooling estimator yields lower loss in the large majority of settings, with the sharpest exceptions with long time series and very high weight on bias.  Moving to the left column, when homogeneity is imposed in the cluster estimator, loss performance relative to simple LP estimation deteriorates, with simple LP estimation dominant in long time series and even short horizons in the upper-left set of horizon-weight combinations.  Reflecting the patterns just described, in the far-right column comparing the LP-pooling estimators with a single vs. multiple clusters, imposing the homogeneity restriction is generally harmful, resulting in higher loss versus the more flexible estimator except with short or very short time series and some long horizons.

\subsection{Coverage and uncertainty quantification}\label{sec:sim-coverage}
Another important aspect of the estimation of impulse response functions is uncertainty quantification, commonly gauged with empirical coverage rates.  Coverage will be affected by not only imprecision in the variance estimate but also bias in the point estimate, a point highlighted in \cite{Oleaetal2025}.  Simple LP estimates report wider intervals than the pooling estimators, as the right column of Figure~\ref{fig:bias-sd} shows. Even so, they tend to yield coverage rates below nominal, more so at longer horizons than shorter \citep{HerbstJohannsen2024, PigerStockwell2025}. 

Accordingly, this subsection examines the empirical coverage of our pooling estimators as compared to simple LP estimators.  We report empirical coverage of the central $1-\alpha=0.90$ interval on data-informed unit-horizon pairs. For the pooling methods we compare the raw posterior credible interval, the classical M\"uller correction with unit-specific $\hat J_{i,h}$, and our proposed correction with cluster-pooled $\hat J_{s,h}$; for naive LP the corresponding object is its own Newey--West interval. HAC (Newey--West at the bandwidth $L_{i,h}$ of Section~\ref{sec:sim-estimators}) is the primary sandwich throughout, consistent with the long-run variance used by the M\"uller correction;  Huber--White (HC) variants lead to the same conclusions.

\begin{figure}[!tbp]\centering
\includegraphics[width=\linewidth]{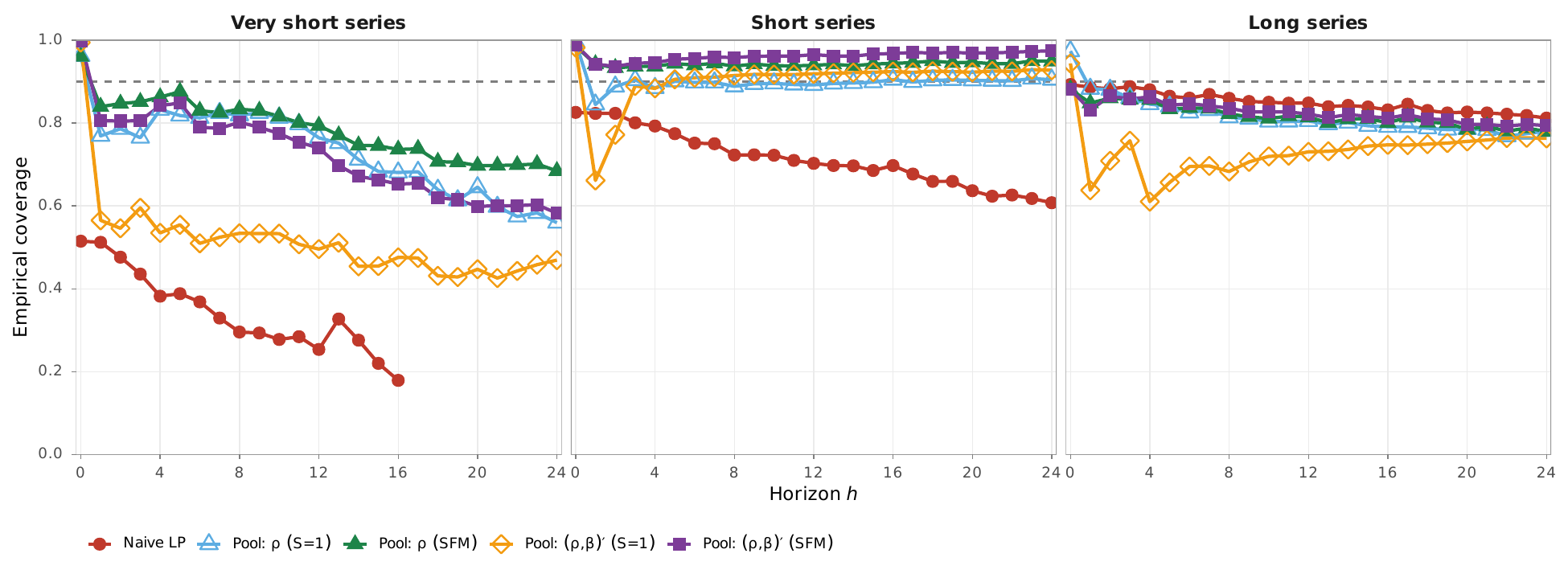}
\caption{Empirical coverage of nominal $0.90$ interval by horizon, FRED-MD calibration.}
\label{fig:coverage-by-h}
\fignotes{Panels, left to right: very short ($T \in [25, 60]$), short ($T \in [100, 150]$), and long ($T = 500$) series.  Dashed grey line: nominal level.  Each method is shown with its headline inference recipe: the unit-specific HAC (Newey--West) sandwich for naive LP and the $S{=}1$ pools, the cluster-pooled HAC sandwich for the SFM methods.  Naive LP is drawn where computable.}
\end{figure}

Table~\ref{tab:coverage} reports our empirical coverage rates, with horizons grouped again as short, medium, and long.  Figure~\ref{fig:coverage-by-h} provides a graphical display covering each horizon.  Once again start with the short time series case.  Consistent with previous studies noted above, the coverage rates of simple LP estimates fall somewhat short of nominal, modestly at short horizons and more substantially at longer horizons.  With the M\"uller correction used, our baseline pooling estimator yields better (in fact slightly conservative) coverage rates, of 95 to 97 percent depending on the horizon.  Among the correction options, the unit-specific $\hat J_{i,h}$ and cluster-pooled $\hat J_{s,h}$ approaches perform comparably.  {The pooling of variances lets the long units of a cluster, whose long-run variances are precisely estimated, dominate the noisier estimates of the shorter units.}  The other pooling specifications that omit pooling of controls' coefficients or impose homogeneity also largely improve on the coverage of the simple LP estimator.

Not surprisingly, with very short time series, accurate coverage becomes more difficult to achieve.  Under-coverage is especially extreme with the simple LP estimator, at roughly 50 percent at short horizons and 25 percent at longer horizons.  Our baseline approach that pools for all coefficients and allows multiple clusters achieves better coverage, not quite at nominal (roughly 85 percent) at shorter horizons and well below nominal (roughly 60 percent) at longer horizons.    Turning to long time series, simple LP estimates yield nearly correct coverage at short and medium horizons, with a little under-coverage at long horizons.  Our proposed pooling estimator achieves nearly the same coverage.  Between the M\"uller correction options, pooling $\hat J_{s,h}$ across clusters comes at little cost relative to using the unit-specific $\hat J_{i,h}$; the coverage rates under these options coincide to within 0.01. 

\input{tables/tab_coverage.tex}

Broadly, this section's Monte Carlo evidence indicates that, in a setting with a panel of related time series including both short and long time samples, our proposed pooling approach to LP estimation can be expected to yield significant benefits in bias, variance, and empirical coverage.  The gains are of course concentrated in the series with short or very short samples.  Our approach that pools across both the LP coefficient and controls' coefficients and allows multiple clusters to be determined by data performs best.  At the same time, especially with our recommended specification there appears to be little cost in bias, variance, and coverage for variables with long time samples.

\section{Empirical application}\label{sec:application}
\subsection{Data, model specification and prior setup}\label{sec:data-spec}

Our application studies how a broad cross section of United States price series responds to two identified structural shocks. The panel of dependent variables comprises $M = 43$ monthly series. Of these, $33$ are consumer and producer price indices, a small number of PCE consumer price series together with a much larger set of producer price indices, and the remaining $10$ are regional Federal Reserve business-survey price measures from the Dallas, Philadelphia, New York, Richmond, and Kansas City Feds, covering manufacturing and services. The emphasis on producer prices is deliberate. Sectoral consumer price responses have already been studied extensively, and those series typically run back to 1959, which weakens the short-sample motivation of our framework. The producer price set instead includes a mixture of longer time series focused on goods and  newer, shorter time series measures with broader coverage of services. These have received far less attention, in part because their shorter samples are difficult to handle with standard methods, and they are the case our pooling approach is designed for.  Similarly, the Federal Reserve measures of prices received include some shorter time series.

We focus on two identified structural shocks. The first is the K\"anzig supply-chain shock \citep{KanzigRaghavan2026}, identified from high-frequency movements in shipping costs around narrative supply-chain events. The second is an oil supply shock, measured by the Baumeister--Hamilton oil supply series \citep{BaumeisterHamilton2019}.\footnote{Both shock series are available on the authors' websites, with the latter shock series regularly updated.} Both shocks operate through the cost of production, which makes the cross section of producer prices a natural place to study their transmission. 

The controls follow choices that are common in the LP and structural VAR literatures on identified shocks in monthly data. For every series we include lags of the consumer price index, the one-year Treasury yield, industrial production, the unemployment rate, the S\&P 500, and the excess bond premium of \citet{GilchristZakrajsek:2012:AER}, using $p = 4$ lags of each. The shock and all remaining regressors are standardized before estimation. When impulse responses are reported, the standardization of the dependent variable is unwound, so that each response reads in that series' own units, while the shock is left on its standardized scale: every impulse response we report is therefore the response to a one-standard-deviation shock.\footnote{The supply-chain shock as constructed in \cite{KanzigRaghavan2026} is normalized to produce (on impact) a 10 percent increase in the real cost of shipping.  As to the oil supply shock of \cite{BaumeisterHamilton2019}, in unreported simple LP estimates, a one standard deviation shock induces an oil price drop of about 2.5 percent on impact, subsequently increasing (in absolute value) over 6 months to about 4 percent.}  We restrict the sample to the pre-COVID period through December 2019, so that no specialized treatment of pandemic-induced high volatility is needed. The shock windows are then 1971:01--2019:12 for the supply-chain shock and 1975:02--2019:12 for the oil supply shock.  Table \ref{tab:appdata} in the appendix lists all the controls and price data series used, along with sources.

For each series and horizon $h$ we regress the $h$-step-ahead price on the current shock $w_t$ and the controls, so that the sequence of shock coefficients $\rho_{i,h}$ over $h$ is the IRF of that series. The $33$ consumer and producer price indices enter as $100 \times \log$ levels, so that each response reads directly in percent.  Among the Fed survey series, all but the Richmond measures are net-balance diffusion indices (for price changes relative to the prior month), which take negative values and so cannot be logged; they enter the LP regressions in their native levels, and their responses read in index points rather than percent.  In the case of the Richmond Fed, the manufacturing and services price series are published as mean percent changes in prices over 12 months rather than net balances.  We iteratively construct a monthly price index from the reported percent changes; {the resulting indices are strictly positive, so they take the same $100 \times \log$ transform and their responses are directly comparable with those of the price indices}.\footnote{Because every unit is standardized before sampling, this mixture of scales causes the pooling no numerical difficulty.} Among the regressors, the shock, the one-year Treasury yield, the unemployment rate, and the excess bond premium enter in levels rather than logs. Alongside the macroeconomic controls we include $p = 4$ lags of the dependent variable itself, and we trace responses out to a maximum horizon of $h = 35$ months, so that the reported IRFs cover $36$ horizons including the impact response. 

We work in levels rather than long differences, following \citet{Oleaetal2025}, who show that differencing the outcome is redundant once a lag of the dependent variable is present.  \citet{PigerStockwell2025} instead recommend the long-difference form, based on evidence of small-sample bias in LP estimates based on levels.  However, in our empirically founded Monte Carlo analysis with data displaying some persistence, our pooled LP estimates showed only small bias.  In keeping with our proposed approach, rather than fitting each variable's LP regression on its own, we estimate them jointly through the hierarchical pooling model of Section~\ref{sec:framework}, so that the shorter producer-price series borrow strength from the longer ones. The resulting IRFs are reported in Section~\ref{sec:app-cluster-irfs}.  Guided by the last section's Monte Carlo results, we focus on our approach that pools across both the LP coefficient and controls' coefficients and allows multiple clusters to be determined by the data, which performs best.

The prior follows the data-calibrated construction of Section~\ref{sec:sim-estimators}, applied to this panel. Every series is standardized per unit before sampling, so the scales below are comparable across units, and the shock is scaled by its own standard deviation. The population center is weakly informative, $m_h \sim \mathcal{N}(0, c \cdot s_y^2)$ with $c = 100$. {The dispersion estimates $\hat{v}_{\text{within}}$ and $\hat{v}_{\text{between}}$ that scale the priors on $\tau_{s,h}^2$ and $B_h^2$ are constructed exactly as in Section~\ref{sec:sim-estimators}, but computed on the long series alone} (those carrying at least $95\%$ of the maximum number of observations, $14$ of the $43$ series in the supply-chain sample and $17$ of $43$ in the oil sample), because on the short series the spread of $\hat\rho_{i,h}$ reflects estimation noise rather than genuine heterogeneity, and calibrating to it would inflate the prior scales and undo the pooling. This gives $b_0 = 0.083$ and $b_B = 0.356$ for the supply-chain shock and $b_0 = 0.063$ and $b_B = 0.431$ for the oil shock. In both samples the calibrated between-cluster variance exceeds the within-cluster variance by a factor of roughly four for the supply-chain shock and roughly seven for the oil shock, which is what makes a mixture worth fitting.

The remaining blocks follow the same specification. We set $S = 8$ components, a generous upper bound that the sparse prior prunes back rather than a choice of the number of clusters. The weights carry a symmetric Dirichlet prior whose concentration is estimated rather than fixed, $e_0 \sim \mathrm{Gamma}(1, 200)$ with prior mean $0.005$ as in the simulation, drawn by the random-walk Metropolis step of Appendix~\ref{app:fullcond}. In the joint pool the deviations of the cluster means from the population center carry a horseshoe with standard half-Cauchy scales on the controls, while the deviations of individual units from their own cluster mean carry a {Gaussian  prior}; the control block is pooled jointly with $\rho$ rather than left unit-specific. The LP residual variances are $\sigma_{i,h}^2 \sim \mathcal{IG}(2.1, b_{\sigma,h})$ with the scale itself estimated, pooled across all series at each horizon, $b_{\sigma,h} \sim \mathrm{Gamma}(1, 2)$. We run the sampler for $10{,}000$ iterations, discard the first $5{,}000$ as burn-in, and retain every subsequent draw. All credible bands reported below apply the cluster-pooled M\"uller correction of Section~\ref{sec:muller}, to the individual-series and the cluster-mean responses alike, so that their widths reflect the serial correlation in the horizon-$h$ LP residual and not the sampler's likelihood alone; the posterior medians are unaffected.

\subsection{Estimates of clusters and impulse responses}\label{sec:app-cluster-irfs}

In discussing the results in this subsection, we will first consider the cluster estimates and then turn to additional detail on other patterns in the responses of more specific price series.

\begin{figure}[htbp]\centering
\includegraphics[width=\linewidth]{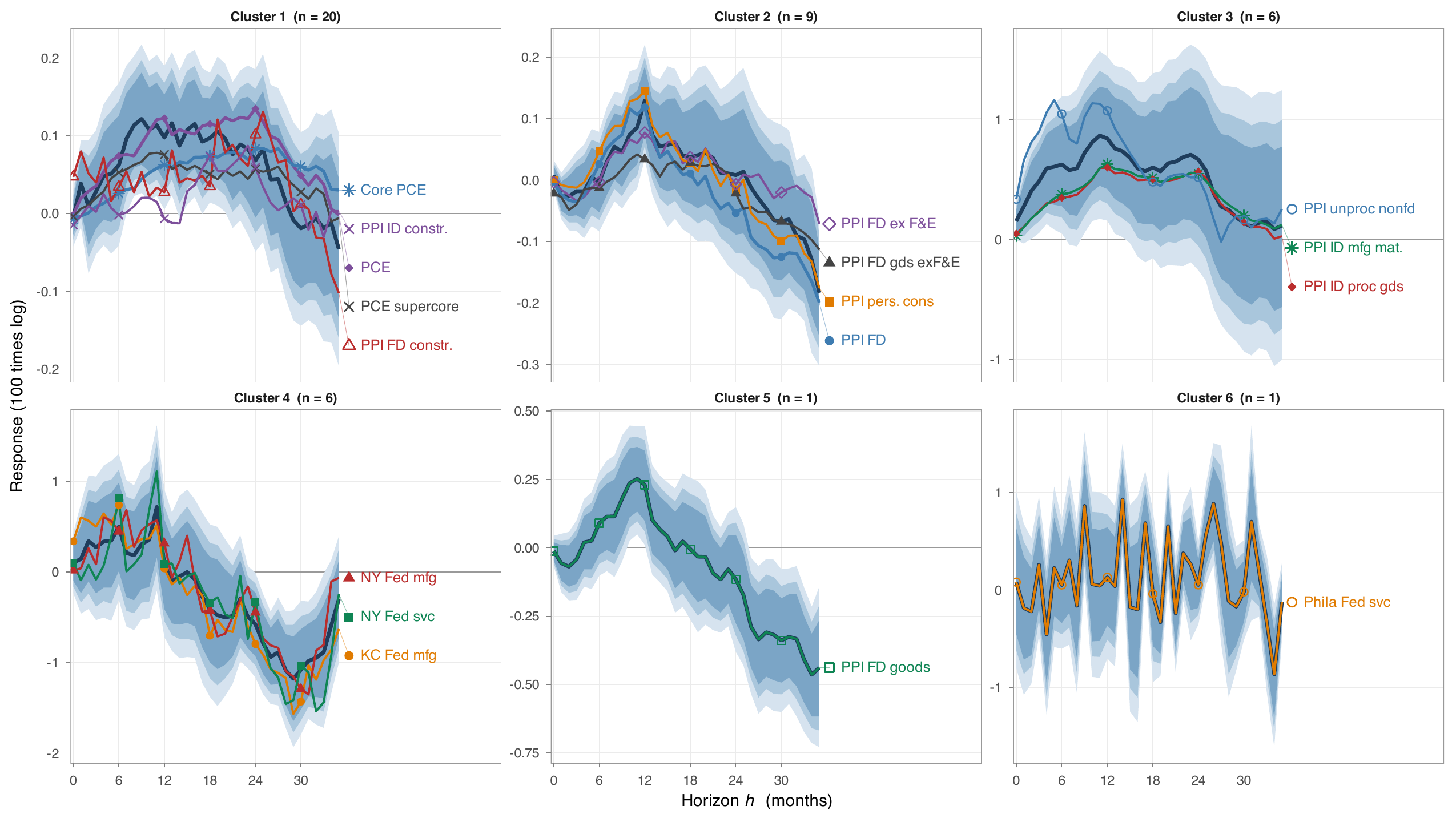}
\caption{Cluster IRFs in response to the K\"anzig supply-chain shock, pre-COVID sample (1971:01--2019:12).}
\label{fig:app-cluster-Kzsupply}
\fignotes{One panel per active mixture cluster (sorted by size). Cluster posterior median (dark line) with 68\%, 80\%, and 90\% pointwise M\"uller-corrected bands (shaded), and posterior median IRFs of selected cluster members (thin lines, each labeled at the right edge with its plotting symbol shown after the name; the same symbol marks the line every six months); the discussion in the text also draws on members that are not shown. All responses are to a one-standard-deviation shock. Units are not uniform across the panel: the $33$ consumer and producer price indices, together with the two Richmond Fed measures (which we cumulate into price levels from their published 12-month mean percent changes), all enter as $100 \times \log$ levels, so their responses read in percent and are mutually comparable. The remaining eight regional-survey series are net-balance diffusion indices entered in their native levels; their responses are in index points and their vertical scale is not comparable with the log-level responses, including where the two kinds share a panel.}
\end{figure}

\begin{figure}[htbp]\centering
\includegraphics[width=\linewidth]{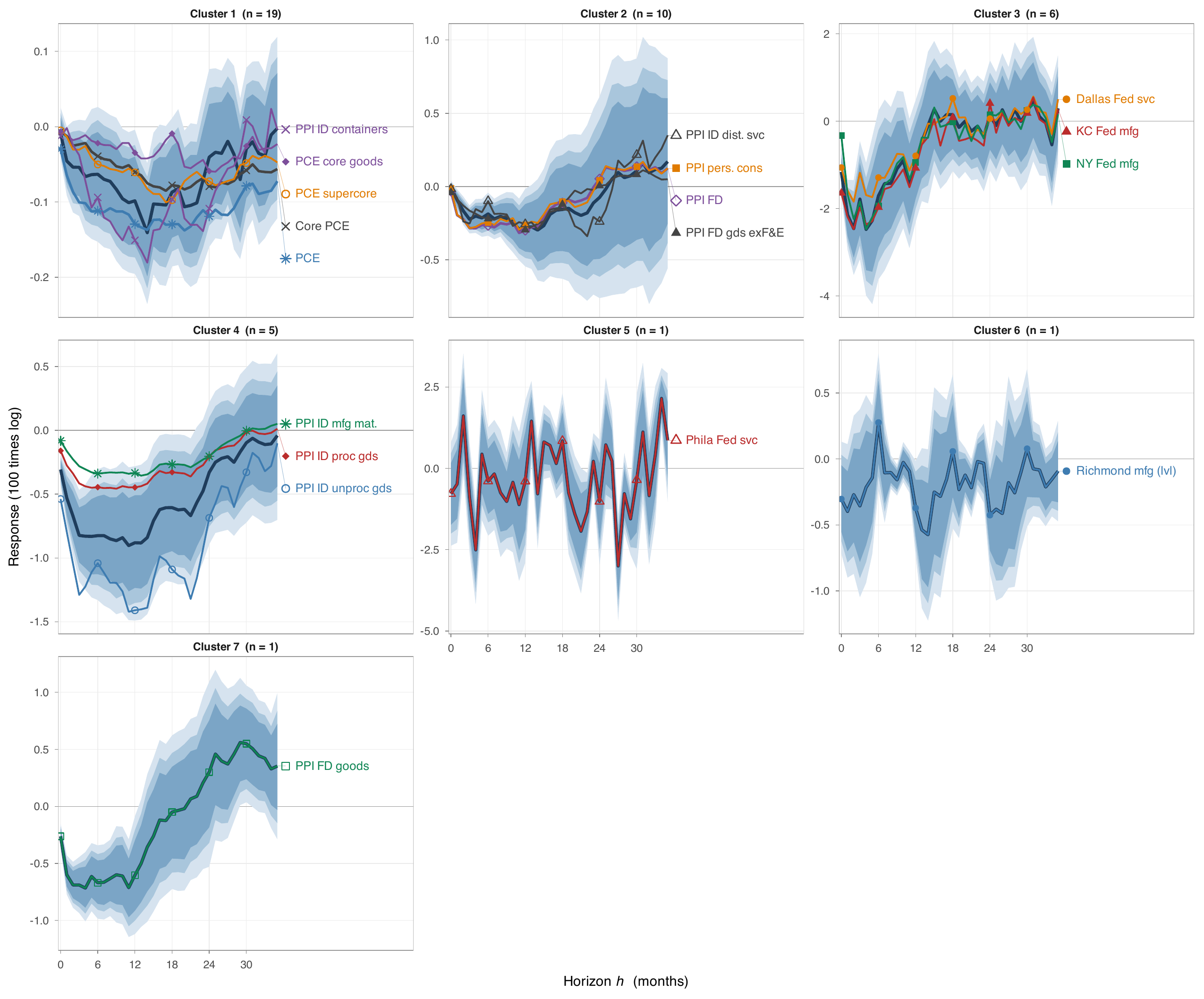}
\caption{Cluster IRFs in response to the Baumeister--Hamilton oil supply shock, pre-COVID sample (1975:02--2019:12).}
\label{fig:app-cluster-BHoilsupply}
\fignotes{Layout as in Figure~\ref{fig:app-cluster-Kzsupply}. Clusters are numbered by size within each shock, so the cluster indices here do not correspond to those of Figure~\ref{fig:app-cluster-Kzsupply}.}
\end{figure}

Table~\ref{tab:clusters} lists the series assigned to each data-determined cluster together with their sample sizes.  The main listing is for the supply-chain shock; the last column gives the cluster to which each variable is assigned in the estimates for the oil supply shock.  Figures~\ref{fig:app-cluster-Kzsupply} and~\ref{fig:app-cluster-BHoilsupply} show the corresponding cluster IRFs from the joint $(\rho, \bm\beta)'$ SFM on the $M = 43$ series, one panel per occupied cluster: the cluster posterior median, $68\%$, $80\%$, and $90\%$ pointwise bands, and the posterior median IRFs of selected member series.  Cluster averages are taken over the modal partition. We should be clear about what the plotted curve is. It is not the hierarchical location $\mu_{s,h}$. That object lives on the standardized scale, and for a cluster whose members are measured differently it has no single representation in original units. What we plot is the average of the members' own responses, $M_s^{-1}\sum_{i: z_i = s}\rho_{i,h}$, each in that series' own units. If the members of a cluster share a scale, the average reads in that scale. If they do not, it is a blend and should be read qualitatively. This definition also explains why panels for singleton clusters display that one series' own posterior rather than a group estimate, so their width reads differently from the rest.

\input{tables/tab_clusters.tex}

From the results on cluster groupings in the table, several  patterns are evident.  First, estimates of responses to the two different shocks yield very similar, but not exactly the same, groupings of series into clusters.  As examples, the series of clusters 1, 2, 3, and 4 are very similarly grouped into clusters for the oil supply shock, although, because clusters are numbered by size within each shock, the cluster indices themselves need not line up across the two sets of estimates.  Second, the estimates indicate that, as intended, clusters pool some short time series with long time series.  For example, with the supply-chain shock, cluster 1 includes many series with more than 500 observations as well as a number of newer PPI measures on construction and services with 122 observations.  Third, the estimated impulse response functions in Figures~\ref{fig:app-cluster-Kzsupply} and~\ref{fig:app-cluster-BHoilsupply} show that, within a cluster, there can be substantial heterogeneity of the series' responses.  For instance, visually, cluster 1 with the supply-chain shock shows clearly visible differences across series, even though these are not quantitatively large, whereas cluster 4 has more sizable quantitative differences across some series.  In addition, some clusters have a single series.  For example, the PPI for final demand goods forms its own cluster.  This is likely connected to magnitudes. This series' response profile looks similar to that of other PPIs in another cluster, but with larger movements.  The Philadelphia Fed services survey likewise forms its own cluster; its response is estimated very imprecisely.  (We return below to additional detail on the responses of goods prices.)

The cluster estimates also show a few notable aspects of groupings of series in clusters.  Among them, cluster 1 is estimated to contain all PCE price indices (aggregate and components), several PPI components of final demand, the PPIs for both final demand and intermediate construction, and some PPIs for final and intermediate demand services.  Among the PPIs for services, most of those for final demand appear in cluster 1, whereas most of those for intermediate demand are assigned to cluster 2.  Most of the PPIs for intermediate demand goods are estimated to belong in a single cluster (cluster 3 for the supply-chain shock and cluster 4 for the oil shock).  Most of the Federal Reserve diffusion indices of prices are also estimated to belong to a single cluster, the exceptions being the two Philadelphia Fed surveys.

\begin{figure}[htbp]\centering
\includegraphics[width=\linewidth]{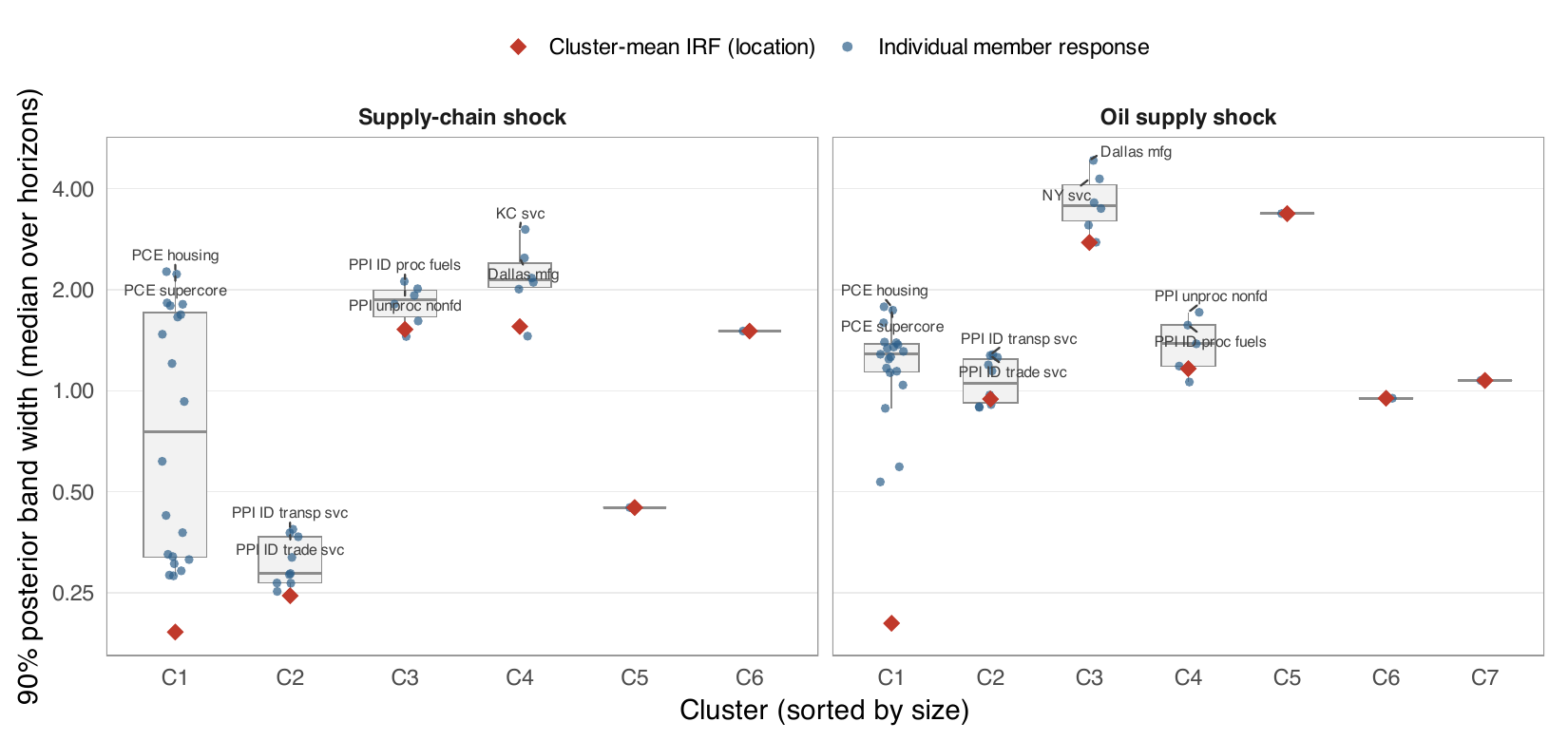}
\caption{Precision of the cluster-mean IRF against the precision of the individual member responses.}
\label{fig:app-precision}
\fignotes{ The vertical axis is on a log scale. The vertical distance from the diamond to the points compares interval widths. It shows how much within-cluster heterogeneity adds to the uncertainty about a single series.  Outlying members are labeled. Widths are on each series' own scale ($100\times\log$ for the price and Richmond series, net-balance for the other survey indices), so the levels are not comparable across scale types; the comparison here is between a cluster mean and its own members, always on a common scale.}
\end{figure}
Figures~\ref{fig:app-cluster-Kzsupply} and~\ref{fig:app-cluster-BHoilsupply} show that, for both shocks, the cluster responses display significant changes in prices for most clusters. The broad consumer-price aggregate (cluster 1) has the tightest band of all.  Its response is away from zero between roughly seven and twenty months for the supply-chain shock, and over the first twenty months for the oil shock.  The other multi-member clusters have wider bands and move away from zero over shorter windows.  Turning to the price dynamics captured in the impulse response estimates, as expected, it is generally the case that the price measures excluding food and energy respond less to shocks than do corresponding headline measures.\footnote{Using CPI measures of consumer prices, \cite{KanzigRaghavan2026} similarly find that food and energy prices respond sharply to a supply-chain shock, but their estimates show headline and core CPI inflation responding similarly.}  In the case of the adverse shock to supply-chain conditions, headline PCE prices post a small rise that peaks after about a year and fades thereafter.  Core PCE prices show a similar contour, but a smaller increase.  Core goods and non-housing services behave similarly.  The housing component shows little change.  Among the PPIs for final demand that cover categories of personal consumption, the impulse response of the PPI for personal consumption services is similar to that for core PCE non-housing services. The same similarity applies with the PPI for finished consumer goods less foods and energy and core PCE goods, but the PPI for consumer goods not omitting food and energy rises more significantly.

Among the PPIs, the overall index for final demand also shows a small rise in response to the supply-chain shock, followed by a decline at longer horizons.  Prices for final demand goods rise somewhat more, but this appears to be driven by the food and energy component; final demand goods less food and energy show little change.  Within goods categories, prices of unprocessed goods for intermediate demand rise sharply, more than those of processed goods for intermediate demand and, in turn, more than those of final demand goods (consistent with \citealp{Clark1999RESTAT}).  Producer prices for services also show little change in response to the supply-chain shock, at both the final and intermediate demand stages.  With services split out by PPI sector, the transportation and warehouse components (both final and intermediate demand) respond more sharply than the others, although the differences are quantitatively small.  The Fed diffusion indices, for both manufacturing and services, are estimated to be supply-chain sensitive, with an initial increase that lasts for about a year followed by a substantial decrease.\footnote{As these are diffusion indices of price changes, these responses mean that, initially, more firms report raising prices than lowering them, but eventually, more firms report lowering prices than increasing them.}

The positive shock to oil supply (which induces a decline in oil prices) yields sectoral impacts that are largely the mirror image of those estimated for the supply-chain shock, with a few notable differences.  With the oil supply shock, the services PPIs for transportation and warehousing respond even more sharply compared to other services components, with a significant decline for the final demand series as well as (a little less sharply) the intermediate series.  With the oil shock, the intermediate demand series for trade services also declines relatively more than other intermediate demand services (the same does not apply for final demand trade services).  The behavior of the Federal Reserve diffusion indices is also somewhat different.  With the oil shock, the diffusion indices decline significantly for about half a year, recover to around zero within roughly a year, and then show little change (whereas they initially rise and later fall in response to the supply-chain shock).

A natural question is whether the cluster bands, which for the large price clusters are visibly tighter than the individual member lines, mean that each individual price series is estimated that precisely. They do not, and the distinction matters.  The comparatively tight bands are for the cluster \emph{location}, the common response its members share, which is estimated from the pooled information of the whole cluster and so sharpens as the cluster acquires more and longer members, provided their estimation errors are not too closely related. An individual series' own response is a different object: it is the cluster location plus that series' idiosyncratic deviation from it, and that deviation carries genuine dispersion that no amount of data about the cluster can remove. 

Figure~\ref{fig:app-precision} makes the gap concrete. For each occupied cluster (sorted by size and across the two shocks), the box and blue points give the $90\%$ band widths (median over horizons) of the individual member series' responses, producing one value per cluster member. The red diamond gives the $90\%$ band width of the cluster-mean IRF, summarized the same way.  In the largest price cluster the cluster-mean band is four to six times tighter than a typical member's own band. The median $90\%$ width is $0.19$ against $0.77$ for the supply-chain shock and $0.20$ against $1.28$ for the oil shock. The gap is much smaller in the other multi-member clusters. In the intermediate-demand goods, final-demand PPI, and the survey group,  the cluster mean is only ten to thirty percent tighter than a typical member. The difference is determined by how closely the members' estimation errors move together. The large price aggregates have largely separate errors, so averaging them cancels a lot of noise. The members of the smaller clusters share most of their residual variation. Averaging then cancels little, and the common location is barely better determined than any single series. For the singleton clusters the two coincide by construction, since the ``cluster'' is the single series itself. The model thus reports high precision only where the data support it, namely for the shared component of a group's response, and does not extend that precision to the individual series.

\section{Conclusions}\label{sec:conclusions}
This paper has developed a Bayesian hierarchical panel model for the estimation of impulse responses to economic shocks. The key idea is to use a sparse finite mixture model to cluster time series of different lengths together and exploit information from longer series to improve estimation of short (and ultra-short) series. In a thorough simulation exercise, we show that this approach pays off. For the short series we focus on, we find substantial improvements in the accuracy of the corresponding local projection-based response estimates. For longer series, our approach produces estimates that are close to the current best-practice benchmark model used in empirical macro. In an application to price series we illustrate that our model recovers economically sensible clusters and show that supply-chain and oil shocks trigger heterogeneous reactions of different price measures, with headline price indices responding more sharply than their core counterparts and goods prices changing more than services prices.

\small{\setstretch{0.75}
\addcontentsline{toc}{section}{References}
\bibliographystyle{frbcle.bst}
\bibliography{lit}}\normalsize\clearpage

\appendix
\newpage

\section{Data}

{
\singlespace
\begin{longtable}{@{}llp{7cm}@{}}
\caption{Application data series and sources:  Controls, prices, and survey measures} \label{tab:appdata} \\
\toprule
\textbf{Series} & \textbf{Source} & \textbf{Description} \\
\midrule
\endfirsthead
\toprule
\textbf{Series} & \textbf{Source} & \textbf{Description} \\
\midrule
\endhead
\midrule
\multicolumn{3}{r}{\textit{Continued on next page}} \\
\endfoot
\bottomrule
\endlastfoot
CPIAUCSL & FRED & CPI \\
GS1 & FRED & 1y Treasury yield \\
INDPRO & FRED & Industrial production \\
UNRATE & FRED & Unemployment rate \\
SP500 & FRED-MD & S\&P 500 \\
EBP & FRED & Excess bond premium \\
PCEPI & FRED & PCE price index \\
PCEPILFE & FRED & Core PCE price index \\
PCEsupercore & BEA website & PCEPI for services excluding energy and housing \\
PCEhousing & BEA website & PCEPI for housing \\
PCEcoregoods & BEA website & PCEPI for goods excluding food and energy \\
PPIFIS & FRED & PPI for FD \\
PPIDGS & FRED & PPI for FD: Goods \\
WPSFD413 & FRED & PPI for FD: FD Goods Less Foods and Energy \\
WPSFD41311 & FRED & PPI for FD: Finished Consumer Goods Less Foods and Energy \\
WPSFD41312 & FRED & PPI for FD: Private Capital Equipment \\
PPIDSS & FRED & PPI for FD: Services \\
PPITSS & FRED & PPI for FD: Trade Services \\
PPIAWS & FRED & PPI for FD: Transportation and Warehousing Services \\
PPITWS & FRED & PPI for FD: Services Less Trade, Transportation, and Warehousing \\
PPIDCS & FRED & PPI for FD: Construction \\
PPIFES & FRED & PPI for FD: Less Foods and Energy \\
WPSFD49501 & FRED & PPI for FD: Personal Consumption \\
WPSFD49502 & FRED & PPI for FD: Personal Consumption Goods \\
WPSFD49505 & FRED & PPI for FD: Personal Consumption Services \\
WPSID61 & FRED & PPI for ID: Processed Goods for ID \\
WPSID611 & FRED & PPI for ID: Materials and Components for Manufacturing \\
WPSID612 & FRED & PPI for ID: Materials and Components for Construction \\
WPSID613 & FRED & PPI for ID: Processed Fuels and Lubricants for ID \\
WPSID614 & FRED & PPI for ID: Containers for ID \\
WPSID615 & FRED & PPI for ID: Supplies for ID \\
WPSID62 & FRED & PPI for ID: Unprocessed Goods for ID \\
WPSID63 & FRED & PPI for ID: Services for ID \\
WPSID631 & FRED & PPI for ID: Services Less Trade, Transportation, and Warehousing for ID \\
WPSID632 & FRED & PPI for ID: Transportation and Warehousing Services for ID \\
WPSID633 & FRED & PPI for ID: Trade Services for ID \\
WPSID64 & FRED & PPI for ID: Construction for ID \\
WPSID69116 & FRED & PPI for ID: Intermediate Distributive Services \\
WPSID69216 & FRED & PPI for ID: Unprocessed Nonfood Materials Less Energy \\
\addlinespace
\multicolumn{3}{@{}l}{\textit{Regional Federal Reserve business-survey price measures}} \\
Dallas\_mfg & Dallas Fed & Manufacturing survey, prices received (net balance) \\
Dallas\_svs & Dallas Fed & Services survey, prices received (net balance) \\
Phil\_mfg & Philadelphia Fed & Manufacturing survey, prices received (net balance) \\
Phil\_svs & Philadelphia Fed & Services survey, prices received (net balance) \\
NY\_mfg & New York Fed & Manufacturing survey, prices received (net balance) \\
NY\_svs & New York Fed & Services survey, prices received (net balance) \\
KC\_mfg & Kansas City Fed & Manufacturing survey, prices received (net balance) \\
KC\_svs & Kansas City Fed & Services survey, prices received (net balance) \\
Rich\_mfg & Richmond Fed & Manufacturing survey, cumulated price level ($100\times\log$) \\
Rich\_svs & Richmond Fed & Services survey, cumulated price level ($100\times\log$) \\
\end{longtable}
Note:  FD and ID refer to final demand and intermediate demand, respectively.
}
\section{Details on posterior simulation}\label{app:fullcond}

This appendix collects the full conditional posterior distributions underlying the Gibbs sampler outlined in Section~\ref{sec:framework} and then sketches the complete algorithm.  Notation follows the main text throughout.  $\bm{Y}_{i,h}$, $\bm{w}_{i,h}$, and $\bm{X}_{i,h}$ denote the stacked LP outcome, structural shock, and controls of series $i$ at horizon $h$, aligned so that each regression uses the $T_{i,h} = \max\{T_i - p - h, 0\}$ observations series $i$ actually provides, and $\bm{W}_{i,h} = (\bm{w}_{i,h}, \bm{X}_{i,h})$ collects the two blocks of regressors of \eqref{eq:lp}.

Two index sets recur: $\mathcal{I}_{s,h} = \{i : z_i = s, H_i \geq h\}$ of \eqref{eq:NW_pooled} collects the units allocated to cluster $s$ that contribute observed information at horizon $h$, and we write $M_{s,h} = |\mathcal{I}_{s,h}|$ for its cardinality, the horizon-$h$ counterpart of the cluster size $M_s$; $\mathcal{S}_h = \{s : \mathcal{I}_{s,h} \neq \emptyset\}$ collects the clusters with at least one such unit, with $S_h^* = |\mathcal{S}_h|$.  As in Section~\ref{sec:framework}, a bar denotes a posterior (here, full conditional) moment, so that $\overline{\rho}_{i,h}$ is the conditional mean of $\rho_{i,h}$ and $\overline{V}_{\rho, i, h}$ its conditional variance, while a hat is reserved for least-squares and other frequentist estimators.

All conditionals are stated for the $\rho$-only pool, in which the mixture prior \eqref{eq: mixture} acts on $\rho_{i,h}$ alone; the modifications under the joint pool of Section~\ref{sec:jointpool}, where the mixture acts on the full coefficient vector $\bm\theta_{i,h} = (\rho_{i,h}, \bm\beta_{i,h}')'$ of dimension $d = 1 + k$, are noted where they arise.  There we index the entries of $\bm\theta_{i,h}$ and of the cluster mean $\bm\mu_{s,h}$ by $j = 0$ for the response and $j = 1, \ldots, k$ for the controls, so that the deviation horseshoe \eqref{eq:level2-beta} governs the controls and $B_h^2$ the response.

A recurring feature is that the hierarchical updates condition only on data-informed quantities: a coefficient at a horizon the series does not reach ($h > H_i$) enters no observation equation, so it integrates out of the relevant full conditional and is re-drawn from its prior at the end of the same iteration.  This is the partially collapsed Gibbs construction of \citet{vanDykPark2008}, and the closing imputation step of the algorithm below is what makes it a valid one.

\subsection{Full conditional posterior sampling}

\paragraph{Impulse responses $\rho_{i,h}$, $h \leq H_i$.}  Because the prior on $\bm\beta_{i,h}$ is informative, we do not marginalize the controls out of the $\rho$-update; instead $(\rho_{i,h}, \bm\beta_{i,h})$ are drawn on an equation-by-equation basis as a Gibbs split.  Given $z_i = s$ and writing $\bm{Y}_{i,h}^{(\beta)} = \bm{Y}_{i,h} - \bm{X}_{i,h}\bm\beta_{i,h}$ for the partial residual at the current draw of $\bm\beta_{i,h}$, we have $\rho_{i,h} \mid \cdot \sim \N\bigl(\overline{\rho}_{i,h}, \overline{V}_{\rho, i, h}\bigr)$ with
\begin{equation}
\overline{V}_{\rho, i, h} = \left(\frac{\bm{w}_{i,h}'\bm{w}_{i,h}}{\sigma_{i,h}^2} + \frac{1}{\tau_{s,h}^2}\right)^{-1}, \qquad
\overline{\rho}_{i,h} = \overline{V}_{\rho, i, h} \left(\frac{\bm{w}_{i,h}'\bm{Y}_{i,h}^{(\beta)}}{\sigma_{i,h}^2} + \frac{\mu_{s,h}}{\tau_{s,h}^2}\right),
\label{eq:rho_post}
\end{equation}
which is the weighted average of Section~\ref{sec:framework}, with the data-based estimate computed on the partial residual: the hierarchical prior enters exactly as one extra, prior-weighted observation.

\paragraph{Control coefficients $\bm\beta_{i,h}$, $h \leq H_i$.}  With prior $\bm{\beta}_{i,h} \sim \N(\bm{0}, \bm{V}_\beta)$ and partial residual $\bm{Y}_{i,h}^{(\rho)} = \bm{Y}_{i,h} - \rho_{i,h} \bm{w}_{i,h}$,
\begin{equation}
\bm{\beta}_{i,h} \mid \cdot \sim \N\bigl(\overline{\bm{\beta}}_{i,h}, \overline{\bm{V}}_{\beta, i, h}\bigr), \quad
\overline{\bm{V}}_{\beta, i, h} = \left(\frac{\bm{X}_{i,h}'\bm{X}_{i,h}}{\sigma_{i,h}^2} + \bm{V}_\beta^{-1}\right)^{-1}, \quad
\overline{\bm{\beta}}_{i,h} = \overline{\bm{V}}_{\beta, i, h} \frac{\bm{X}_{i,h}'\bm{Y}_{i,h}^{(\rho)}}{\sigma_{i,h}^2}.
\label{eq:beta_post}
\end{equation}
The prior covariance $\bm{V}_\beta$ is either fixed and weakly informative or a horseshoe scale-mixture, in which case the local and global scales are updated within the same Gibbs cycle through the auxiliary inverse-gamma representation of \citet{makalic2015simple}.  Alternating \eqref{eq:rho_post} and \eqref{eq:beta_post} targets the exact joint posterior of $(\rho_{i,h}, \bm\beta_{i,h}')'$.

\paragraph{Joint pool: one block for $\bm\theta_{i,h}$, $h \leq H_i$.}  When the controls are pooled alongside the response (Section~\ref{sec:jointpool}), the Gibbs split is replaced by a single Gaussian block given the cluster mean $\bm\mu_{s,h}$ and within-cluster variance $\tau_{s,h}^2$:
\begin{equation}
\bm\theta_{i,h} \mid \cdot \sim \N\left(\overline{\bm{V}}_{\theta, i, h}\left[\frac{\bm{W}_{i,h}'\bm{Y}_{i,h}}{\sigma_{i,h}^2} + \frac{\bm\mu_{s,h}}{\tau_{s,h}^2}\right], \overline{\bm{V}}_{\theta, i, h}\right), \quad
\overline{\bm{V}}_{\theta, i, h} = \left(\frac{\bm{W}_{i,h}'\bm{W}_{i,h}}{\sigma_{i,h}^2} + \frac{\mathbf{I}_d}{\tau_{s,h}^2}\right)^{-1},
\label{eq:theta_post}
\end{equation}
which automatically respects the within-cluster prior \eqref{eq:jointpool} on the response and on every control.

\paragraph{Imputation of $\rho_{i,h}$ for $h > H_i$.}  At horizons a series is too short to reach, the likelihood is empty and the conditional collapses to the within-cluster prior:
\begin{equation}
\rho_{i,h} \mid \cdot \sim \N\bigl(\mu_{z_i,h}, \tau_{z_i,h}^2\bigr), \qquad \text{joint pool:} \quad \bm\theta_{i,h} \mid \cdot \sim \N\bigl(\bm\mu_{z_i,h}, \tau_{z_i,h}^2 \mathbf{I}_d\bigr).
\label{eq:rho_imputation}
\end{equation}
These draws are predictive realizations from the model: they equip short series with IRFs beyond their sample length together with fully propagated uncertainty.

\paragraph{Residual variances $\sigma_{i,h}^2$.}  The scale of the inverse Gamma prior is itself estimated and pooled across series, $\sigma_{i,h}^2 \sim \mathcal{IG}(a_\sigma, b_{\sigma, h})$, with a common horizon-specific scale $b_{\sigma, h} \sim \mathrm{Gamma}(c_\sigma, d_\sigma)$.  Because the scale carries no cluster index, the prior on $\sigma_{i,h}^2$ does not depend on the allocation $z_i$, which is what keeps the allocation conditional \eqref{eq:z_post} below exact.  With residual $\hat{\bm{u}}_{i,h} = \bm{Y}_{i,h} - \rho_{i,h} \bm{w}_{i,h} - \bm{X}_{i,h}\bm\beta_{i,h}$ evaluated at the current draws,
\begin{equation}
\sigma_{i,h}^2 \mid \cdot \sim \mathcal{IG}\left(a_\sigma + \frac{T_{i,h}}{2}, b_{\sigma, h} + \frac{1}{2}\hat{\bm{u}}_{i,h}'\hat{\bm{u}}_{i,h}\right),
\label{eq:sigma_post}
\end{equation}
while the pooled scale is drawn from its own conjugate Gamma conditional over the units with data at horizon $h$,
\begin{equation}
b_{\sigma, h} \mid \cdot \sim \mathrm{Gamma}\left(c_\sigma + M_h a_\sigma, d_\sigma + \sum_{i : H_i \geq h} \sigma_{i,h}^{-2}\right), \qquad M_h = \#\{i : H_i \geq h\}.
\label{eq:bsigma_post}
\end{equation}
No update of $\sigma_{i,h}^2$ is performed for $h > H_i$, where the regression has no observations.

\paragraph{Cluster means $\mu_{s,h}$.}  Conditional on the data-informed members of cluster $s$, $\mu_{s,h} \mid \cdot \sim \N\bigl(\overline{\mu}_{s,h}, \overline{V}_{\mu, s, h}\bigr)$ with
\begin{equation}
\overline{V}_{\mu, s, h} = \left(\frac{M_{s,h}}{\tau_{s,h}^2} + \frac{1}{B_h^2}\right)^{-1}, \qquad
\overline{\mu}_{s,h} = \overline{V}_{\mu, s, h} \left(\frac{1}{\tau_{s,h}^2}\sum_{i \in \mathcal{I}_{s,h}} \rho_{i,h} + \frac{m_h}{B_h^2}\right).
\label{eq:mu_post}
\end{equation}
If $M_{s,h} = 0$, the conditional reduces to the level-2 prior \eqref{eq:level2}, $\N(m_h, B_h^2)$.  Members of cluster $s$ that do not reach horizon $h$ do not enter the update: their imputed coefficients enter no observation equation at horizon $h$, so they integrate out of this full conditional.  Under the joint pool the update applies entry by entry to $\bm\mu_{s,h}$, with $\rho_{i,h}$ replaced by entry $j$ of $\bm\theta_{i,h}$, $m_h$ by $m_{h,j}$, and the level-2 prior variance $B_h^2$ replaced by $\psi_j^2 \psi_B^2$ for the controls $j = 1, \ldots, k$, per the deviation horseshoe \eqref{eq:level2-beta}.

\paragraph{Within-cluster variances $\tau_{s,h}^2$.}  With prior $\tau_{s,h}^2 \sim \mathcal{IG}(a_0, b_0)$,
\begin{equation}
\tau_{s,h}^2 \mid \cdot \sim \mathcal{IG}\left(a_0 + \frac{M_{s,h}}{2}, b_0 + \frac{1}{2}\sum_{i \in \mathcal{I}_{s,h}}\bigl(\rho_{i,h} - \mu_{s,h}\bigr)^2\right),
\label{eq:tau_post}
\end{equation}
reducing to the prior for empty clusters.  Under the joint pool the single scale governs all $d$ entries, so the shape parameter becomes $a_0 + M_{s,h} d / 2$ and the sum of squares runs over every entry of $\bm\theta_{i,h} - \bm\mu_{s,h}$.

\paragraph{Population center $m_h$.}  Conditional on the cluster means of the occupied clusters, and with the level-3 prior \eqref{eq:level3},
\begin{equation}
m_h \mid \cdot \sim \N\left(\frac{\overline{V}_{m, h}}{B_h^2}\sum_{s \in \mathcal{S}_h} \mu_{s,h}, \overline{V}_{m, h}\right), \qquad
\overline{V}_{m, h} = \left(\frac{S_h^*}{B_h^2} + \frac{1}{c \cdot s_y^2}\right)^{-1}.
\label{eq:barmu_post}
\end{equation}
Under the joint pool the same update applies entry by entry to the vector $\bm m_h$ with entries $m_{h,j}$, each carrying the mildly informative prior $\N(0, c \cdot s_y^2)$; for the controls $j = 1, \ldots, k$ the level-2 variance $B_h^2$ in \eqref{eq:barmu_post} is replaced by $\psi_j^2 \psi_B^2$, per the deviation horseshoe \eqref{eq:level2-beta}.  The local scale $\psi_j$ is indexed per control and shared across clusters and horizons, so the model makes a single heterogeneity decision per control rather than one per (cluster, horizon, control) triple.  The scales $(\psi_1^2, \ldots, \psi_k^2, \psi_B^2)$ are drawn once per iteration from the auxiliary inverse-gamma conditionals of \citet{makalic2015simple}, pooling the deviations $\mu_{s,h,j} - m_{h,j}$ of the occupied cluster means across clusters and horizons; deviations of empty clusters are prior draws and integrate out by the same argument.

\paragraph{Between-cluster variances $B_h^2$.}  With prior $B_h^2 \sim \mathcal{IG}(a_B, b_B)$,
\begin{equation}
B_h^2 \mid \cdot \sim \mathcal{IG}\left(a_B + \frac{S_h^*}{2}, b_B + \frac{1}{2}\sum_{s \in \mathcal{S}_h}\bigl(\mu_{s,h} - m_h\bigr)^2\right).
\label{eq:B_post}
\end{equation}
Under the joint pool, $B_h^2$ governs the response only (the sum runs over the entries $\mu_{s,h,0}$ of the occupied cluster means), while the controls carry the deviation-horseshoe scales of \eqref{eq:level2-beta}.

\paragraph{Cluster allocations $z_i$.}  The imputed coefficients at horizons $h > H_i$ integrate out before $z_i$ is drawn, leaving a product over the data-informed horizons only:
\begin{equation}
\mathrm{Prob}(z_i = s \mid \cdot) \propto \pi_s \prod_{h \leq H_i} \phi\bigl(\rho_{i,h} \mid \mu_{s,h}, \tau_{s,h}^2\bigr),
\label{eq:z_post}
\end{equation}
where $\phi(\cdot \mid \mu, \sigma^2)$ denotes the Gaussian density; $z_i$ is drawn from the categorical distribution obtained by normalizing over $s = 1, \ldots, S$.  Under the joint pool each horizon contributes $\prod_{j=0}^{k} \phi(\theta_{i,h,j} \mid \mu_{s,h,j}, \tau_{s,h}^2)$.  Note that the allocation of series $i$ is informed by all of its data-informed horizons jointly, so that a series is grouped by the shape of its whole IRF profile rather than by its response at any single horizon.

\paragraph{Mixture weights $\bm\pi$.}  With $M_s = \#\{i : z_i = s\}$ denoting the number of series in cluster $s$, as in Section~\ref{sec:framework},
\begin{equation}
\bm{\pi} \mid \cdot \sim \mathrm{Dir}(e_0 + M_1, \ldots, e_0 + M_S).
\label{eq:w_post}
\end{equation}

\paragraph{Concentration parameter $e_0$.}  This is the only non-conjugate step.  With prior $e_0 \sim \mathrm{Gamma}(a_e, b_e)$, the log conditional is
\begin{equation}
\log p(e_0 \mid \bm{\pi}) \propto (a_e - 1)\log e_0 - b_e e_0 + \log \Gamma(S e_0) - S \log \Gamma(e_0) + (e_0 - 1) \sum_{s=1}^{S} \log \pi_s,
\label{eq:e0_post}
\end{equation}
and, following \citet{malsinerwalli2016}, we sample $e_0$ by a short Metropolis--Hastings step with a random-walk proposal on $\log e_0$, tuned towards an acceptance rate of roughly $30\%$.

\subsection{Sketch of the algorithm}\label{app:gibbs}

The sampler is initialized with equation-by-equation OLS estimates for the unit-level coefficients, a $K$-means allocation of the standardized series for $z_i$, and prior means for the hierarchical parameters.  Each iteration first applies a random permutation of the cluster labels, so that the sampler explores the full label-symmetric posterior (Section~\ref{sec:framework}), and then cycles through the following blocks:
\begin{enumerate}[leftmargin=*, itemsep=2pt, topsep=2pt]
\item \textbf{Unit-level coefficients and residual variances.}  For each horizon $h$ and each unit $i$ with $H_i \geq h$: under the $\rho$-only pool, alternate the conditionals \eqref{eq:rho_post} and \eqref{eq:beta_post}; under the joint pool, draw $\bm\theta_{i,h}$ in one block from \eqref{eq:theta_post}.  Then draw $\sigma_{i,h}^2$ from \eqref{eq:sigma_post} and the horizon-level scales $b_{\sigma, h}$ from \eqref{eq:bsigma_post}.
\item \textbf{Cluster-level and population-level parameters.}  For each horizon $h$: draw $\mu_{s,h}$ from \eqref{eq:mu_post} and $\tau_{s,h}^2$ from \eqref{eq:tau_post} for every cluster, then $m_h$ from \eqref{eq:barmu_post} (entry by entry under the joint pool) and $B_h^2$ from \eqref{eq:B_post}.  Under the joint pool, close the block by updating the deviation-horseshoe scales $(\psi_j^2, \psi_B^2)$ of \eqref{eq:level2-beta}.
\item \textbf{Allocations, weights, and concentration.}  Draw each $z_i$ from \eqref{eq:z_post}, the weights $\bm\pi$ from \eqref{eq:w_post}, and $e_0$ by the Metropolis--Hastings step on \eqref{eq:e0_post}.
\item \textbf{Imputation.}  For every unit $i$ and every horizon $h > H_i$, re-draw $\rho_{i,h}$ (under the joint pool, $\bm\theta_{i,h}$) from the newly assigned cluster's prior \eqref{eq:rho_imputation}.
\end{enumerate}
Placing the imputation step at the end of the iteration, after the allocation step, is what makes the collapsed conditionals \eqref{eq:mu_post}, \eqref{eq:tau_post}, and \eqref{eq:z_post} valid Gibbs steps: the marginalized coefficients are re-drawn from their level-1 prior before any subsequent step conditions on them, which is the ordering rule of \citet{vanDykPark2008}.  It also gives the chain proper mixing on the imputed coefficients, whose draws always reflect the current cluster parameters and allocation.  Finally, the M\"{u}ller correction of Section~\ref{sec:muller} is a post-processing pass over the saved draws and leaves every step above unchanged.

\end{document}

%% file: tables/tab_mae_headline.tex
\begin{table}[h!]\centering\footnotesize
\caption{Headline accuracy: MAE relative to naive LP on data-informed unit-horizon pairs, FRED-MD calibrated DGP (500 MC replications).}\label{tab:mae-headline}
\setlength{\tabcolsep}{3pt}
\begin{tabular}{l|ccc|ccc|ccc}\toprule
& \multicolumn{3}{c|}{Very short ($T \in [25, 60]$)}
& \multicolumn{3}{c|}{Short ($T \in [100, 150]$)}
& \multicolumn{3}{c}{Long ($T = 500$)}\\
Horizon $h$ (months) & $[0,4]$ & $[5,12]$ & $[13,24]$ & $[0,4]$ & $[5,12]$ & $[13,24]$ & $[0,4]$ & $[5,12]$ & $[13,24]$\\
\midrule
\rowcolor{gray!15}
Naive LP (raw MAE)                         & 0.375$^{\dagger}$ & 0.510$^{\dagger}$ & 0.766$^{\dagger}$ & 0.114 & 0.149 & 0.170 & 0.041 & \textbf{0.055} & \textbf{0.065}\\
Pool: $\rho$ ($S{=}1$)                     & 0.40$^{***}$ & 0.22$^{***}$ & 0.12$^{***}$ & 0.70$^{***}$ & 0.66$^{***}$ & 0.53$^{***}$ & \textbf{0.93}$^{***}$ & 1.10$^{***}$ & 1.14$^{***}$\\
Pool: $\rho$ (SFM)                         & 0.34$^{***}$ & 0.19$^{***}$ & \textbf{0.10}$^{***}$ & 0.58$^{***}$ & 0.57$^{***}$ & 0.53$^{***}$ & 1.07$^{***}$ & 1.07$^{***}$ & 1.07$^{***}$\\
Pool: $(\rho,\bm\beta)'$ ($S{=}1$)         & 0.34$^{***}$ & 0.22$^{***}$ & 0.12$^{***}$ & 0.93$^{***}$ & 0.64$^{***}$ & \textbf{0.46}$^{***}$ & 1.60$^{***}$ & 1.42$^{***}$ & 1.24$^{***}$\\
Pool: $(\rho,\bm\beta)'$ (SFM)             & \textbf{0.24}$^{***}$ & \textbf{0.16}$^{***}$ & \textbf{0.10}$^{***}$ & \textbf{0.55}$^{***}$ & \textbf{0.51}$^{***}$ & 0.47$^{***}$ & 1.07$^{***}$ & 1.06$^{***}$ & 1.05$^{***}$\\
\bottomrule
\end{tabular}
\fignotes{The second header row gives the horizon buckets $h$ in months.  All entries refer to MAE on data-informed unit-horizon pairs ($h \le H_i$).  The shaded naive LP row reports raw (absolute) MAE; every other row is expressed relative to naive LP (values below one indicate higher accuracy; a method's absolute MAE is the product of its entry and the naive row).  Bold marks the most accurate method per column, ties at the displayed precision sharing the bold.  $^{*}$/$^{**}$/$^{***}$: the method's MAE differs from naive LP's at the $10$/$5$/$1\%$ level (two-sided paired $t$-test of the per-replication MAE difference across the $500$ replications).  All estimators condition on the same information set.  $^{\dagger}$MAE over the $19\%$ of data-informed pairs where naive LP is computable (footnote~\ref{fn:naive-veryshort}).}
\end{table}

%% file: tables/tab_coverage.tex
\begin{table}[h!]\centering\footnotesize
\caption{Empirical coverage of the nominal $0.90$ interval on data-informed unit-horizon pairs, FRED-MD calibrated DGP (500 MC replications).}\label{tab:coverage}
\setlength{\tabcolsep}{2.2pt}
\begin{tabular}{ll|ccc|ccc|ccc}\toprule
& & \multicolumn{3}{c|}{Very short ($T \in [25, 60]$)}
& \multicolumn{3}{c|}{Short ($T \in [100, 150]$)}
& \multicolumn{3}{c}{Long ($T = 500$)}\\
Method & Interval & $[0,4]$ & $[5,12]$ & $[13,24]$ & $[0,4]$ & $[5,12]$ & $[13,24]$ & $[0,4]$ & $[5,12]$ & $[13,24]$\\
\midrule
Naive LP                                     & HAC, unit                & 0.47 & 0.32 & 0.25 & 0.81 & 0.73 & 0.66 & 0.89 & 0.86 & 0.83 \\
\midrule
Pool: $\rho$ ($S{=}1$)                       & credible                 & 0.77 & 0.75 & 0.64 & 0.80 & 0.72 & 0.66 & 0.85 & 0.74 & 0.63 \\
                                             & HAC, unit                & 0.82 & 0.81 & 0.65 & 0.90 & 0.89 & 0.90 & 0.89 & 0.82 & 0.79 \\
\midrule
Pool: $\rho$ (SFM)                           & credible                 & 0.67 & 0.70 & 0.71 & 0.53 & 0.47 & 0.46 & 0.63 & 0.53 & 0.49 \\
                                             & HAC, unit                & 0.83 & 0.81 & 0.72 & 0.94 & 0.92 & 0.91 & 0.87 & 0.83 & 0.81 \\
                                             & HAC, pooled              & 0.87 & 0.83 & 0.72 & 0.95 & 0.94 & 0.94 & 0.86 & 0.82 & 0.80 \\
\midrule
Pool: $(\rho,\bm\beta)'$ ($S{=}1$)           & credible                 & 0.48 & 0.37 & 0.44 & 0.56 & 0.49 & 0.62 & 0.63 & 0.47 & 0.52 \\
                                             & HAC, unit                & 0.65 & 0.52 & 0.46 & 0.84 & 0.91 & 0.92 & 0.73 & 0.70 & 0.75 \\
\midrule
Pool: $(\rho,\bm\beta)'$ (SFM)               & credible                 & 0.60 & 0.63 & 0.62 & 0.46 & 0.44 & 0.46 & 0.53 & 0.47 & 0.45 \\
                                             & HAC, unit                & 0.83 & 0.76 & 0.63 & 0.94 & 0.94 & 0.94 & 0.86 & 0.84 & 0.82 \\
                                             & HAC, pooled              & 0.85 & 0.79 & 0.63 & 0.95 & 0.96 & 0.97 & 0.86 & 0.83 & 0.81 \\
\bottomrule
\end{tabular}
\fignotes{The second header row gives the horizon buckets $h$ in months, as in Table~\ref{tab:mae-headline}.  ``credible'' is the standard posterior credible interval; ``HAC, unit'' is the M\"uller correction with unit-specific $\hat J_{i,h}$ (for naive LP, its own Newey--West interval); ``HAC, pooled'' uses the observation-weighted cluster-pooled $\hat J_{s,h}$ of \eqref{eq:NW_pooled}.  In the very-short block naive LP covers only the pairs where it is computable (footnote~\ref{fn:naive-veryshort}).}
\end{table}

%% file: tables/tab_clusters.tex
\begin{table}[!ht]\centering\scriptsize
\caption{Estimated partition of the 43 price and survey series, with each series' effective sample length.}\label{tab:clusters}
\setlength{\tabcolsep}{3pt}
\scalebox{0.82}{%
\begin{tabular}[t]{@{}>{\raggedright\arraybackslash}p{5.6cm}lrrc@{}}
\toprule
Series & Mnemonic & $T_i$ & $T_{i,H}$ & Oil cl. \\
\midrule
\multicolumn{5}{@{}l}{\textit{Cluster 1} ($M_1 = 20$)} \\
\quad Core PCE & \texttt{PCEPILFE} & 588 & 549 & 1 \\
\quad PCE (headline) & \texttt{PCEPI} & 588 & 549 & 1 \\
\quad PCE goods excl. food/energy & \texttt{PCEcoregoods} & 588 & 549 & 1 \\
\quad PCE housing & \texttt{PCEhousing} & 588 & 549 & 1 \\
\quad PCE services excl. energy/housing & \texttt{PCEsupercore} & 588 & 549 & 1 \\
\quad Philadelphia Fed mfg survey, prices & \texttt{Phil\_mfg} & 588 & 549 & 1 \\
\quad PPI FD: personal consumption goods & \texttt{WPSFD49502} & 588 & 549 & 1 \\
\quad PPI ID: construction materials & \texttt{WPSID612} & 564 & 525 & 1 \\
\quad PPI ID: containers & \texttt{WPSID614} & 564 & 525 & 1 \\
\quad PPI FD: finished cons. goods less food/energy & \texttt{WPSFD41311} & 552 & 513 & 1 \\
\quad PPI ID: supplies & \texttt{WPSID615} & 528 & 489 & 1 \\
\quad Richmond Fed services survey, price level & \texttt{Rich\_svs} & 291 & 252 & 1 \\
\quad Richmond Fed mfg survey, price level & \texttt{Rich\_mfg} & 255 & 216 & 6 \\
\quad PPI FD: construction & \texttt{PPIDCS} & 122 & 83 & 1 \\
\quad PPI FD: services & \texttt{PPIDSS} & 122 & 83 & 1 \\
\quad PPI FD: services less trade/transp. & \texttt{PPITWS} & 122 & 83 & 1 \\
\quad PPI FD: trade services & \texttt{PPITSS} & 122 & 83 & 1 \\
\quad PPI ID: construction & \texttt{WPSID64} & 122 & 83 & 1 \\
\quad PPI ID: services & \texttt{WPSID63} & 122 & 83 & 2 \\
\quad PPI ID: services less trade/transp. & \texttt{WPSID631} & 122 & 83 & 1 \\
\addlinespace
\multicolumn{5}{@{}l}{\textit{Cluster 2} ($M_2 = 9$)} \\
\quad PPI FD: goods less food/energy & \texttt{WPSFD413} & 122 & 83 & 2 \\
\quad PPI FD: transp./warehousing svcs & \texttt{PPIAWS} & 122 & 83 & 2 \\
\quad PPI final demand (headline) & \texttt{PPIFIS} & 122 & 83 & 2 \\
\quad PPI ID: trade services & \texttt{WPSID633} & 122 & 83 & 2 \\
\quad PPI ID: transp./warehousing svcs & \texttt{WPSID632} & 122 & 83 & 2 \\
\quad PPI FD: personal consumption & \texttt{WPSFD49501} & 120 & 81 & 2 \\
\quad PPI FD: less food/energy & \texttt{PPIFES} & 117 & 78 & 2 \\
\quad PPI FD: personal consumption services & \texttt{WPSFD49505} & 117 & 78 & 2 \\
\quad PPI ID: distributive services & \texttt{WPSID69116} & 117 & 78 & 2 \\
\bottomrule
\end{tabular}
\hspace{1.2em}%
\begin{tabular}[t]{@{}>{\raggedright\arraybackslash}p{5.6cm}lrrc@{}}
\toprule
Series & Mnemonic & $T_i$ & $T_{i,H}$ & Oil cl. \\
\midrule
\multicolumn{5}{@{}l}{\textit{Cluster 3} ($M_3 = 6$)} \\
\quad PPI FD: private capital equipment & \texttt{WPSFD41312} & 588 & 549 & 1 \\
\quad PPI ID: processed goods & \texttt{WPSID61} & 588 & 549 & 4 \\
\quad PPI ID: unprocessed goods & \texttt{WPSID62} & 588 & 549 & 4 \\
\quad PPI ID: mfg materials & \texttt{WPSID611} & 564 & 525 & 4 \\
\quad PPI ID: processed fuels/lubricants & \texttt{WPSID613} & 564 & 525 & 4 \\
\quad PPI ID: unproc. nonfood materials less energy & \texttt{WPSID69216} & 552 & 513 & 4 \\
\addlinespace
\multicolumn{5}{@{}l}{\textit{Cluster 4} ($M_4 = 6$)} \\
\quad Kansas City Fed mfg survey, prices & \texttt{KC\_mfg} & 222 & 183 & 3 \\
\quad New York Fed mfg survey, prices & \texttt{NY\_mfg} & 222 & 183 & 3 \\
\quad Dallas Fed mfg survey, prices & \texttt{Dallas\_mfg} & 187 & 148 & 3 \\
\quad New York Fed services survey, prices & \texttt{NY\_svs} & 184 & 145 & 3 \\
\quad Dallas Fed services survey, prices & \texttt{Dallas\_svs} & 156 & 117 & 3 \\
\quad Kansas City Fed services survey, prices & \texttt{KC\_svs} & 72 & 33 & 3 \\
\addlinespace
\multicolumn{5}{@{}l}{\textit{Cluster 5} ($M_5 = 1$)} \\
\quad PPI FD: goods & \texttt{PPIDGS} & 122 & 83 & 7 \\
\addlinespace
\multicolumn{5}{@{}l}{\textit{Cluster 6} ($M_6 = 1$)} \\
\quad Philadelphia Fed services survey, prices & \texttt{Phil\_svs} & 106 & 67 & 5 \\
\bottomrule
\end{tabular}
}
\fignotes{Clusters are the modal allocations under the supply-chain shock (sample 1971:01--2019:12), sorted by size; ``Oil cl.'' gives the modal cluster under the oil supply shock (sample 1975:02--2019:12), whose partition has 7 rather than 6 groups. $T_i$ is the number of monthly observations the series contributes and $T_{i,H} = T_i - p - H$ the number still available at the longest horizon shown, which ranges from 33 for the shortest series to 549 for the longest: no series exhausts its own sample within $H$, but the short ones become very thin, and it is there that pooling sharpens them most. The 33 price indices and the two Richmond Fed measures, which we cumulate into price levels from their published 12-month mean percent changes, enter as $100 \times \log$ levels; the other eight regional-survey series are net-balance diffusion indices entered in their native levels. For the oil sample every $T_i$ is shorter by up to 49 months, since that shock starts later.}
\end{table}